\documentclass[epj]{svjour}

\usepackage[utf8]{inputenc}
\usepackage[T1]{fontenc}
\usepackage{amsmath,amsfonts,amssymb}
\usepackage{graphicx}
\usepackage{subcaption}
\usepackage{xcolor}
\usepackage{booktabs}
\usepackage[colorlinks=true,
            linkcolor=blue,
            citecolor=blue,
            urlcolor=blue]{hyperref}

\graphicspath{{Images/}}

\begin{document}

\title{Equilibrium Halo Solutions of the Gross--Pitaevskii--Poisson System: The Role of the Particle Number}

\author{
Francisco A. Guzm\'an\inst{1}
\and
El\'ias Castellanos\inst{2}
\and
Jorge Mastache\inst{3,4}
}

\institute{
Facultad de Ciencias en F\'isica y Matem\'aticas,
Universidad Aut\'onoma de Chiapas,
Tuxtla Guti\'errez, Chiapas, M\'exico
\and
Escuela de Ingenier\'ia y Ciencias,
Tecnol\'ogico de Monterrey,
Carr. Lago de Guadalupe Km. 3.5,
Estado de M\'exico 52926, M\'exico\\
\email{elias.castellanos@tec.mx}
\and
Secretar\'ia de Ciencia, Humanidades, Tecnolog\'ia e Innovaci\'on (SECIHTI),
Av. Insurgentes Sur 1582, Colonia Cr\'edito Constructor,
Ciudad de M\'exico, 03940, M\'exico
\and
Mesoamerican Centre for Theoretical Physics,
Universidad Aut\'onoma de Chiapas,
Carretera Zapata Km. 4, Real del Bosque,
29040, Tuxtla Guti\'errez, Chiapas, M\'exico\\
\email{jh.mastache@mctp.mx}
}

\abstract{
    We investigate stationary halo-like solutions of the Gross--Pitaevskii--Poisson (GPP) system describing self-gravitating Bose--Einstein condensates with repulsive self-interactions as a model for dark matter. We retain explicitly the boson mass $m_\phi$, the scattering length $a_s$, and the total particle number $N$, treating the latter as an independent macroscopic control parameter of the equilibrium configuration. We solve the stationary GPP equations and explore a broad region of parameter space. The resulting equilibrium configurations are classified into ground-state, excited-state, and unbound branches according to their binding properties and nodal structure. We find that the ground-state branch occupies a well-defined region of the $(m_\phi,N)$ plane whose location depends sensitively on the self-interaction strength, while the excited-state and unbound regions exhibit a structure largely independent of the initial ansatz. By analyzing the converged solutions, we derive empirical scaling relations linking the characteristic halo radius $R_{99}$ to the boson mass, scattering length, and total particle number. The recovered scaling reproduces the known mass--radius relation in the non-interacting limit, while finite self-interactions reveal an intermediate regime in which gravity, quantum pressure, and repulsive interactions jointly determine the equilibrium structure, deviating from the asymptotic Thomas--Fermi limit. As an astrophysical application, we show that ground-state solutions can reproduce the rotation curves of representative dwarf galaxies using only the solitonic component. Finally, we examine the impact of repulsive self-interactions on the location of the ground-state branch and discuss the implications for current Lyman-$\alpha$ forest constraints. Although increasing $a_s$ shifts equilibrium solutions toward boson masses compatible with existing Lyman-$\alpha$ bounds, the corresponding halo configurations fail to reproduce the observed dwarf-galaxy kinematics. Our results provide a systematic characterization of stationary GPP halos and establish a direct connection between the fundamental particle properties $(m_\phi,a_s)$, the macroscopic control parameter $N$, and observable galactic properties.
}

\maketitle

\section{Introduction}

The possibility that dark matter is composed of ultralight bosons has motivated extensive studies of self-gravitating quantum fluids described by the Gross--Pitaevskii--Poisson (GPP), or equivalently Schr\"odinger--Poisson, system. In the non-relativistic mean-field regime, a Bose--Einstein condensate (BEC) evolving under its own gravitational field is described by a macroscopic wavefunction, or order parameter, coupled self-consistently to the Newtonian gravitational potential. This framework has been widely explored in fuzzy, ultralight, scalar-field, and wave dark matter models \cite{Hui:2016ltb,Schive:2014hza,Suarez:2013iw,Matos:1999et,Matos:2000jx,Urena-Lopez:2001zjo,Robles:2012uy,Guzman:2004wj,Guzman:2006yc,Sin:1992bg,Lee:1995af,Peebles:2000yy,Goodman:2000tg,Arbey:2003sj,Matos:2000ng,Harko:2011xw,Marsh:2015xka,Hui:2016ltb}.

The physical appeal of this scenario is that a large occupation number allows the dark matter distribution to be modeled as a coherent field rather than as an ensemble of individual particles. In this description, the density is determined by the squared amplitude of the condensate wavefunction, while quantum pressure, self-interactions, and gravity jointly determine the equilibrium structure \cite{Pitaevskii2016,Pethick:2008,Hui:2016ltb,Chavanis:2011zi,Widrow:1993qq,Madelung:1927ksh,Chavanis:2011zi,Suarez:2013iw,Hui:2016ltb}. For a condensate with short-range interactions, the GPP system takes the form
\begin{align}
    i\hbar \frac{\partial \Psi}{\partial t} &= \left( -\frac{\hbar^2}{2m_{\phi}}\nabla^2 + m_{\phi}\Phi + g_s|\Psi|^2 \right)\Psi, \label{eq:gp_intro}. \\
    \nabla^2\Phi &= 4\pi Gm_{\phi}|\Psi|^2, \label{eq:poisson_intro}
\end{align}
where $m_{\phi}$ is the boson mass, $g_s=4\pi\hbar^2a_s/m_{\phi}$, $a_s$ is the $s$-wave scattering length, and $\Psi$ is normalized to the total particle number $N$.

Cosmological simulations of ultralight dark matter indicate that halos develop a central solitonic core surrounded by an extended interference-supported envelope \cite{Schive:2014hza,Hui:2016ltb,Mocz:2017wlg,Mocz:2019pyf,Schive:2014dra,Nori:2018pka}. The solitonic core corresponds to the ground-state configuration of the Schr\"odinger--Poisson/GPP system and arises from the balance between self-gravity and quantum pressure, the latter originating from the kinetic-energy term of the wavefunction (or equivalently from the Bohm quantum potential in the Madelung fluid representation), which opposes gravitational compression on small scales. In the presence of repulsive self-interactions, the Thomas--Fermi limit provides another analytically tractable regime, where self-interaction pressure also contributes to supporting the configuration \cite{Boehmer:2007um,Chavanis:2011zi,Chavanis:2011uv,Colpi:1986ye,Harko:2011xw,Chavanis:2011zm}. These ideas have been applied to galactic halos and rotation curves in both phenomenological and dynamical contexts \cite{Boehmer:2007um,Robles:2012uy,Castellanos:2019ttq}.

The equilibrium properties of self-gravitating bosonic systems also have a long history in the boson-star literature. Relativistic boson stars were first studied in Refs.~\cite{Kaup:1968zz,Ruffini:1969qy}, while self-interacting configurations were shown to admit substantially larger masses in Ref.~\cite{Colpi:1986ye}. In the Newtonian regime, stationary and time-dependent Schr\"odinger--Poisson simulations have shown that generic configurations can relax toward ground-state solitons through gravitational cooling, while excited branches with radial nodes also exist as stationary solutions of the nonlinear eigenvalue problem \cite{Seidel:1993zk,Guzman:2004wj,Guzman:2006yc,Schunck:2003kk,Liebling:2012fv,Chavanis:2025qcg}.

Despite this progress, most studies either focus on dynamical formation, exploit scaling symmetries to generate families of solutions, or use phenomenological halo profiles to fit observations \cite{Guzman:2004wj,Guzman:2006yc,Schive:2014hza,Chavanis:2011zi,Hui:2016ltb,Mocz:2017wlg,Castellanos:2020omh}. The direct construction of finite-domain, halo-like equilibrium configurations from the stationary GPP equations remains comparatively less explored, particularly when the fundamental particle properties $(m_{\phi},a_s)$ and the total particle number $N$ are kept explicit and is treated as an independent physical control parameter.

This point is important because $N$ fixes the total halo mass, $M_{\rm H}=m_{\phi}N$, and enters both the normalization of the wavefunction and the outer boundary condition of the gravitational potential. Therefore, keeping $N$ explicit provides a direct map between microscopic particle parameters and macroscopic halo observables such as $R_{99}$, the density profile, and the circular velocity curve. While equilibrium sequences parametrized by conserved quantities are standard in the boson-star literature \cite{Kaup:1968zz,Ruffini:1969qy,Colpi:1986ye,Schunck:2003kk}, here we apply this viewpoint to halo-motivated GPP solutions in a finite radial domain.

In this work we construct spherically symmetric stationary solutions of the GPP system using a finite-difference discretization and a Newton--Raphson iterative scheme. The solver is initialized with several finite-mass ans\"atze---Gaussian, exponential, linear--exponential, and hyperbolic-secant profiles---in order to test the dependence of convergence and solution classification on the initial profile. The resulting configurations are classified as ground states, excited states, or unbound configurations according to their binding energy and nodal structure.

We then map the $(m_{\phi},N)$ solution space for different values of the scattering length, extract empirical scaling relations for the characteristic radius $R_{99}$, and compute the circular velocity profiles of the converged ground-state configurations. This allows us to connect stationary GPP equilibria directly with galactic rotation-curve observables.

Finally, we examine the relevance of these solutions in light of Lyman-$\alpha$ forest constraints. Current analyses generally disfavor the canonical fuzzy-dark-matter mass scale $m_{\phi}\sim10^{-22}\,\mathrm{eV}$ when ultralight bosons make up all of the dark matter, with lower bounds ranging from the $10^{-21}\,\mathrm{eV}$ scale to $m_{\phi}>2\times10^{-20}\,\mathrm{eV}$, depending on the data set and modeling assumptions \cite{Irsic:2017yje,Armengaud:2017nkf,Rogers:2020ltq,Eberhardt:2025caq}. We therefore investigate whether repulsive self-interactions shift the ground-state region toward masses compatible with these bounds and whether such configurations can still reproduce dwarf-galaxy rotation curves.

The main contribution of this paper is the explicit construction and characterization of stationary, halo-like GPP equilibrium solutions with physical parameters kept explicit. This provides a controlled framework to study how $m_{\phi}$, $N$, and $a_s$ determine halo size, density profiles, solution branches, and rotation curves.

\section{The Gross--Pitaevskii--Poisson system}\label{sec:gpp_system}

We consider a self-gravitating Bose--Einstein condensate (BEC) composed of identical bosons of mass $m_{\phi}$ in the non-relativistic mean-field regime. In this approximation, the many-body system is described by a macroscopic wavefunction, or order parameter, whose evolution is governed by the Gross--Pitaevskii (GP) equation. The GP formalism provides the standard mean-field description of dilute Bose gases and has been widely applied to self-gravitating condensates in the context of ultralight dark matter and boson-star models \cite{Pitaevskii2016,Pethick:2008,Chavanis:2011zi,Hui:2016ltb}.

\subsection{Stationary formulation}

The dynamics of the condensate are described by the Gross--Pitaevskii equation
\begin{equation} \label{eq:GP_dependiente} 
    i\hbar\frac{\partial\Psi}{\partial t} = \left( -\frac{\hbar^2}{2m_{\phi}}\nabla^2 +V +g_s|\Psi|^2 \right)\Psi,
\end{equation}
where
\begin{equation}
    g_s=\frac{4\pi\hbar^2a_s}{m_{\phi}}
\end{equation}
is the coupling constant associated with the $s$-wave scattering length $a_s$, and $V$ denotes an external potential.

The macroscopic wavefunction is normalized to the total number of particles,
\begin{equation} \label{NormalizationN}
    N = \int |\Psi(\mathbf r,t)|^2,d^3r.
\end{equation}

We seek stationary configurations of the form
\begin{equation}\label{Psi_space_time}
    \Psi(\mathbf r,t) = e^{-i\mu t/\hbar}\psi(\mathbf r),
\end{equation}
where $\mu$ is the chemical potential and $\psi(\mathbf r)$ is a time-independent order parameter satisfying the same normalization condition,
\begin{equation}
    N = \int |\psi(\mathbf r)|^2,d^3r.
\end{equation}

Substituting Eq.\eqref{Psi_space_time} into Eq.\eqref{eq:GP_dependiente} yields the stationary Gross--Pitaevskii equation
\begin{equation} \label{Eq:Stationary_GP}
    \mu\psi = -\frac{\hbar^2}{2m_{\phi}}\nabla^2\psi + V\psi + g_s|\psi|^2\psi.
\end{equation}

The Gross--Pitaevskii--Poisson (GPP) system is obtained by identifying the external potential with the self-generated gravitational potential, $V=m_{\phi}\Phi$, where $\Phi$ satisfies the Newtonian Poisson equation
\begin{equation} \label{Poisson_general}
    \nabla^2\Phi = 4\pi G m_{\phi}|\psi|^2.
\end{equation}

Equations~\eqref{Eq:Stationary_GP} and \eqref{Poisson_general} define a nonlinear eigenvalue problem for the pair $(\psi,\mu)$. The chemical potential $\mu$ plays the role of a nonlinear eigenvalue, while the condensate profile $\psi$ is the associated eigenfunction.

It is important to emphasize that Eq.\eqref{Eq:Stationary_GP} is not restricted to ground-state configurations. This can be understood already in the linear limit. If the self-interaction is neglected and the potential is regarded as fixed, Eq.\eqref{Eq:Stationary_GP} reduces to the stationary Schr\"odinger equation,
\begin{equation}
    \hat H \psi_j = \left( -\frac{\hbar^2}{2m_{\phi}}\nabla^2 + V \right)\psi_j = \mu_j\psi_j,
\end{equation}
whose bound solutions form a discrete spectrum of eigenfunctions.

In the spherically symmetric case, these eigenfunctions are naturally ordered by their nodal structure: the lowest-energy state is nodeless, while higher-energy states possess one or more radial nodes. Although the Gross--Pitaevskii equation is nonlinear and the superposition principle no longer applies, stationary solutions can still be organized into branches characterized by the number of nodes of the order parameter \cite{Guzman:2004wj,Guzman:2006yc,Chavanis:2011zi}.

Throughout this work we classify nodeless solutions as ground states, while solutions exhibiting one or more radial nodes are classified as excited states. Although the density
\begin{equation}
    \rho(r) = m_{\phi}|\psi(r)|^2,
\end{equation}
remains positive, the presence of nodes indicates that the condensate belongs to a higher-energy stationary branch. Time-dependent studies of the Schrodinger--Poisson and GPP systems have shown that excited states generally relax toward the ground-state branch through gravitational cooling, supporting the interpretation of the nodeless solution as the physically preferred equilibrium configuration \cite{Guzman:2004wj,Guzman:2006yc}.

\subsection{Variational structure}

The stationary GPP equations can be derived from a variational principle. Equilibrium configurations correspond to extrema of the energy functional at fixed particle number,
$\delta(E-\mu N)=0$, where $\mu$ appears as a Lagrange multiplier enforcing the normalization constraint \cite{Pitaevskii2016,Pethick:2008,Chavanis:2011zi}.

For a self-gravitating condensate, the total energy is
\begin{equation}\label{Energy_system}
    E[\psi] = \int d^3r \left[ \frac{\hbar^2}{2m_{\phi}}|\nabla\psi|^2 + \frac{g_s}{2}|\psi|^4 + \frac{1}{2}m_{\phi}\Phi|\psi|^2 \right].
\end{equation}
The first term corresponds to the kinetic energy, the second term to the short-range self-interaction energy, and the third term to the gravitational interaction energy. The factor $1/2$ in the latter avoids double counting of pairwise gravitational interactions.

Along a sequence of equilibrium configurations, the chemical potential satisfies
\begin{equation}\label{mu_fracc} 
    \mu = \frac{\partial E}{\partial N}.
\end{equation}

Multiplying Eq.\eqref{Eq:Stationary_GP} by $\psi^{*}$ and integrating over the volume yields
\begin{equation}\label{mu}
    \mu N = \int \left[ -\frac{\hbar^2}{2m_{\phi}} \psi^{*}\nabla^2\psi + m_{\phi}\Phi|\psi|^2 + g_s|\psi|^4 \right] d^3r.
\end{equation}

\subsection{Spherically symmetric GPP system}

We focus on isolated halo-like configurations and assume spherical symmetry, $\psi=\psi(r)$, and $\qquad \Phi=\Phi(r)$. The stationary GPP equation becomes
\begin{equation} \label{radial_GP}
    \mu\psi = -\frac{\hbar^2}{2m_{\phi}} \left( \frac{d^2\psi}{dr^2} + \frac{2}{r}\frac{d\psi}{dr} \right) + m_{\phi}\Phi\psi + g_s|\psi|^2\psi,
\end{equation}
while the Poisson equation reduces to
\begin{equation} \label{radial_Poisson} 
    \frac{1}{r^2} \frac{d}{dr} \left( r^2\frac{d\Phi}{dr} \right) = 4\pi G m_{\phi}|\psi|^2.
\end{equation}

Regularity at the origin requires
\begin{equation}
\left.\frac{d\psi}{dr}\right|_{r=0}=0,
\qquad
\left.\frac{d\Phi}{dr}\right|_{r=0}=0.
\end{equation}

Equations\eqref{radial_GP}--\eqref{radial_Poisson} define a coupled nonlinear eigenvalue problem for $(\psi,\Phi)$, with $\mu$ determined by the normalization constraint.

The enclosed mass inside radius $r$ is
\begin{equation} \label{MASAENCERRADAr}
    M_{\mathrm{enc}}(r) = 4\pi \int_0^r r'^2\rho(r'),dr'.
\end{equation}
Equation~\eqref{NormalizationN} implies that the total halo mass is $M_{\mathrm H} = m_{\phi}N$.

Using Gauss's law, the gravitational field can be written as
\begin{equation}
\frac{d\Phi}{dr}
=
\frac{G M_{\mathrm{enc}}(r)}{r^2},
\end{equation}
showing explicitly that the gravitational potential is determined by the cumulative mass distribution.

Unlike many studies that exploit the scaling symmetries of the GPP system and work exclusively with dimensionless variables, we retain the total particle number $N$ as an explicit physical parameter throughout the analysis. Consequently, the fundamental particle properties $(m_{\phi},a_s)$ together with the macroscopic control parameter $N$ determine the properties of the equilibrium configuration, providing a direct connection between particle physics parameters and observable halo quantities.

\subsection{Dimensionless formulation and scaling symmetries}

The Gross--Pitaevskii--Poisson system possesses well-known scaling symmetries that are frequently exploited to construct families of solutions from a single dimensionless configuration \cite{Chavanis:2011zi,Guzman:2004wj,Hui:2016ltb}. In particular, in the non-interacting limit ($a_s=0$), the stationary GPP equations remain invariant under a rescaling of the spatial coordinates, density, and gravitational potential, implying that a single numerical solution can generate an entire family of equilibrium configurations through an appropriate change of scale.

For this reason, many studies formulate the problem in terms of dimensionless variables, solving the equations in rescaled units and recovering physical quantities only at the end of the calculation. This approach is particularly useful when studying generic properties of the solution space, such as mass--radius relations or stability criteria.

In the present work, however, we adopt a complementary strategy and retain the particle properties $(m_{\phi},a_s)$ and the total particle number $N$ explicitly throughout the calculation. The total particle number $N$ is treated as an independent control parameter rather than being absorbed into a scaling transformation. As a result, each numerical solution corresponds directly to a physical halo configuration with total mass $M_{\rm H}=m_{\phi}N$. This choice provides a transparent connection between microscopic particle properties and macroscopic halo observables, facilitating the interpretation of the parameter-space exploration presented in Sec.~\ref{sec:results}.

\section{Numerical method}\label{sec:numerical_method}

We solve the stationary Gross--Pitaevskii--Poisson system under spherical symmetry using a finite-difference discretization of the radial domain combined with a Newton--Raphson iterative scheme. Similar finite-difference approaches have been widely employed in the study of stationary Schr\"odinger--Poisson and boson-star systems \cite{Guzman:2004wj,Guzman:2006yc,Schunck:2003kk}. The present implementation is adapted to the construction of halo-like equilibrium configurations while keeping the particle properties $(m_{\phi},a_s)$ and the total particle number $N$ explicit throughout the calculation.

\subsection{Radial discretization}
Assuming spherical symmetry, the equations reduce to a system of
ordinary differential equations on the interval $r\in[\Delta r,R]$, where $R$
is chosen sufficiently large to contain the mass of the configuration.
The domain is discretized into $n$ grid points:
\begin{equation}
    r_i = (i+1)\,\Delta r, \qquad i=0,\dots,n-1,
\end{equation}
with uniform spacing $\Delta r = R/n$. 

Second-order centered finite-difference stencils are used to approximate both first- and second-order radial derivatives. The discretization transforms the stationary GPP equations into a coupled system of sparse algebraic equations defined on the radial mesh. The origin requires special treatment in order to avoid divergences at $r=0$, therefore, the first point of the interval is taken to be $\Delta r$, which satisfies the condition $\Delta r \to 0$ as $n$ increases.

\subsection{Newton--Raphson linearization}

The stationary Gross--Pitaevskii equation can be written as a nonlinear
operator equation
\begin{equation}
    L(\psi) = 0,
\end{equation}
with
\begin{equation}
    L(\psi) = \mu \psi + \frac{\hbar^{2}}{2m_{\phi}}\nabla^{2}\psi - m_{\phi}\Phi\,\psi - g_{s}|\psi|^{2}\psi .
\end{equation}

Because of the cubic self-interaction term, the stationary GPP system defines a nonlinear eigenvalue problem. We solve it iteratively using a Newton--Raphson scheme, which linearizes the equations around a trial solution and successively improves the approximation until convergence is achieved. Expanding $L(\psi)$ around an approximate solution $\psi^{(k)}$ gives
\begin{equation}
    L(\psi^{(k)} + \delta\psi) \approx L(\psi^{(k)}) + L'(\psi^{(k)})\,\delta\psi ,
\end{equation}
where higher-order terms are neglected.

Imposing $L(\psi^{(k+1)})=0$ leads to the linear correction equation
\begin{equation}\label{N_Lineal}
    L'(\psi^{(k)})\,\delta\psi^{(k)} = L(\psi^{(k)}),
\end{equation}
with update rule
\begin{equation}\label{correcion_k}
    \psi^{(k+1)} = \psi^{(k)} - \delta\psi^{(k)} .
\end{equation}

The linearized operator is
\begin{equation}\label{NpsiDer}
    L'(\psi^{(k)}) = \mu^{(k)} + \frac{\hbar^{2}}{2m_{\phi}}\nabla^{2} - m_{\phi}\Phi^{(k)} - 3g_{s}|\psi^{(k)}|^{2}.
\end{equation}
At each iteration the correction equation is discretized into a linear system that can be solved efficiently using standard matrix methods. The resulting matrices are sparse and retain the banded structure associated with the finite-difference representation of the radial Laplacian, making the iterative procedure computationally efficient.

\subsection{Self-consistent update of the potential}

After each update of the wavefunction, the gravitational potential is
recomputed by solving the discretized Poisson equation
\begin{equation}
    \nabla^2 \Phi^{(k+1)} = 4\pi G m_{\phi} |\psi^{(k+1)}|^2 .
\end{equation}
The chemical potential is updated using Eq.~\eqref{mu}, and the wavefunction is renormalized to satisfy Eq.~\eqref{NormalizationN}. This ensures conservation of the total particle number throughout the iteration.

The procedure is repeated until convergence is achieved, defined by
\begin{equation}
    \max_i |\psi^{(k+1)}_i - \psi^{(k)}_i|
    < \varepsilon ,
\end{equation}
with tolerance $\varepsilon$ typically chosen between $10^{-8}$ and $10^{-10}$.

\subsection{Initial profiles (ans\"atze)}

The Newton--Raphson method requires an initial guess sufficiently close
to the true solution. We consider several physically motivated
finite-mass profiles:
\begin{align}
    \psi(r) &\sim \exp{\left(-\frac{r^2}{2r_{d}^2}\right)} \quad \text{\footnotesize Gaussian,} \label{Gaussian_ans} \\
    \psi(r) &\sim \exp{\left(-\frac{r}{r_{d}}\right)} \quad \text{\footnotesize Exponential,} \label{Exp_ans}\\
    \psi(r) &\sim \left(1 + \frac{r}{r_{d}}\right)\exp{\left(-\frac{r}{r_{d}}\right)} \quad \text{\footnotesize Linear+Exp,} \label{Linexp_ans}\\
    \psi(r) &\sim \operatorname{sech}\left(\frac{r}{r_{d}}\right) \quad \text{\footnotesize Hyperbolic Secant.} \label{Sech_ans}
\end{align}
Here $r_d$ denotes a characteristic scale radius controlling the concentration of the initial profile. Since each ansatz possesses a different radial dependence, the fraction of enclosed mass within a given multiple of $r_d$ is profile dependent. The corresponding enclosed mass
fraction
\begin{equation}
    f_M(r) = \frac{M_{\mathrm{enc}}(r)}{M_{\mathrm{H}}}
\end{equation}
is used to determine a suitable outer radius $R$. For $R/r_d \gtrsim 5$ the enclosed mass is essentially constant for all profiles, ensuring that the finite computational domain contains the entire configuration. The use of multiple ans\"atze allows us to assess the robustness of the numerical method and to investigate the sensitivity of the convergence properties and solution classification to the choice of initial profile.

\subsection{Boundary conditions}
For the gravitational potential we impose regularity at the origin and
the asymptotic Newtonian form at the outer boundary,
\begin{align}
    \frac{d\Phi}{dr}\Big|_{r=0} &= 0, \\
    \Phi(R) &= -\frac{G M_{\mathrm{H}}}{R},
\end{align}
which corresponds to the exterior Newtonian potential generated by the total enclosed mass $M_{\mathrm H}$.

For the wavefunction we require regularity at the origin and vanishing radial flux at the boundary,
\begin{equation}
    \frac{d\psi}{dr}\Big|_{r=0} = 0, \qquad \frac{d\psi}{dr}\Big|_{r=R} = 0 .
\end{equation}
These conditions are consistent with a localized bound state inside a finite computational domain.

\subsection{Physical constraints}
To ensure that the solutions correspond to self-gravitating bound
configurations, we impose the condition
\begin{equation} \label{bound_condition}
    \frac{E}{N} < m_{\phi} \Phi(R) ,
\end{equation}
where $m_{\phi}\Phi$ is the potential energy. This condition ensures that the mean energy per particle remains below the gravitational potential energy at the outer boundary, indicating a gravitationally bound configuration.

The nodal structure of the converged solution is recorded as part of the classification procedure introduced in Sec.~2. Solutions belonging to the ground-state branch and to excited-state branches are both retained during the parameter-space exploration, while configurations that fail the binding criterion Eq.~\eqref{bound_condition} are classified as unbound.

\section{Results}\label{sec:results}
In this section we explore the structure of the stationary solution space of the GPP system as a function of the particle properties $(m_{\phi},a_s)$ and the total particle number $N$. Particular emphasis is placed on the role of the total particle number $N$ as an explicit control parameter, the existence of ground-state and excited-state branches, and the connection between stationary GPP solutions and observable halo properties. 

We explore the parameter space $(m_{\phi},N,a_s)$, while fixing the scattering length $a_s$ in the mass range of $m \in [10^{-23}, 10^{-20}]\,\mathrm{eV}$ and $N \in [10^{93}, 10^{99}]\,$, dividing each interval in $40$ equal parts resulting in a $40\times40$ mesh. For each point in parameter space, the stationary GPP equations are solved numerically within a finite radial domain of size $R=100\,\mathrm{kpc}$ using the numerical method described in Sec.~\ref{sec:numerical_method}. The resulting configurations are classified according to the criteria introduced in Sec.~\ref{sec:gpp_system}. Solutions satisfying the binding condition Eq.\eqref{bound_condition} are separated into ground-state and excited-state branches according to the nodal structure of the order parameter, while configurations that fail the binding criterion are classified as unbound.

For each converged solution we compute the total halo mass, $M_{\mathrm H}=m_{\phi}N,$ and the characteristic radius $R_{99}$ enclosing $99\%$ of the total mass. The quantity $R_{99}$ provides a physically meaningful measure of the spatial extent of the halo and will be used throughout this section to characterize the dependence of the solutions on the physical parameters of the configuration.

\begin{figure*}[t]
\centering
\begin{subfigure}{0.45\textwidth}
\centering
    \includegraphics[width=\linewidth]{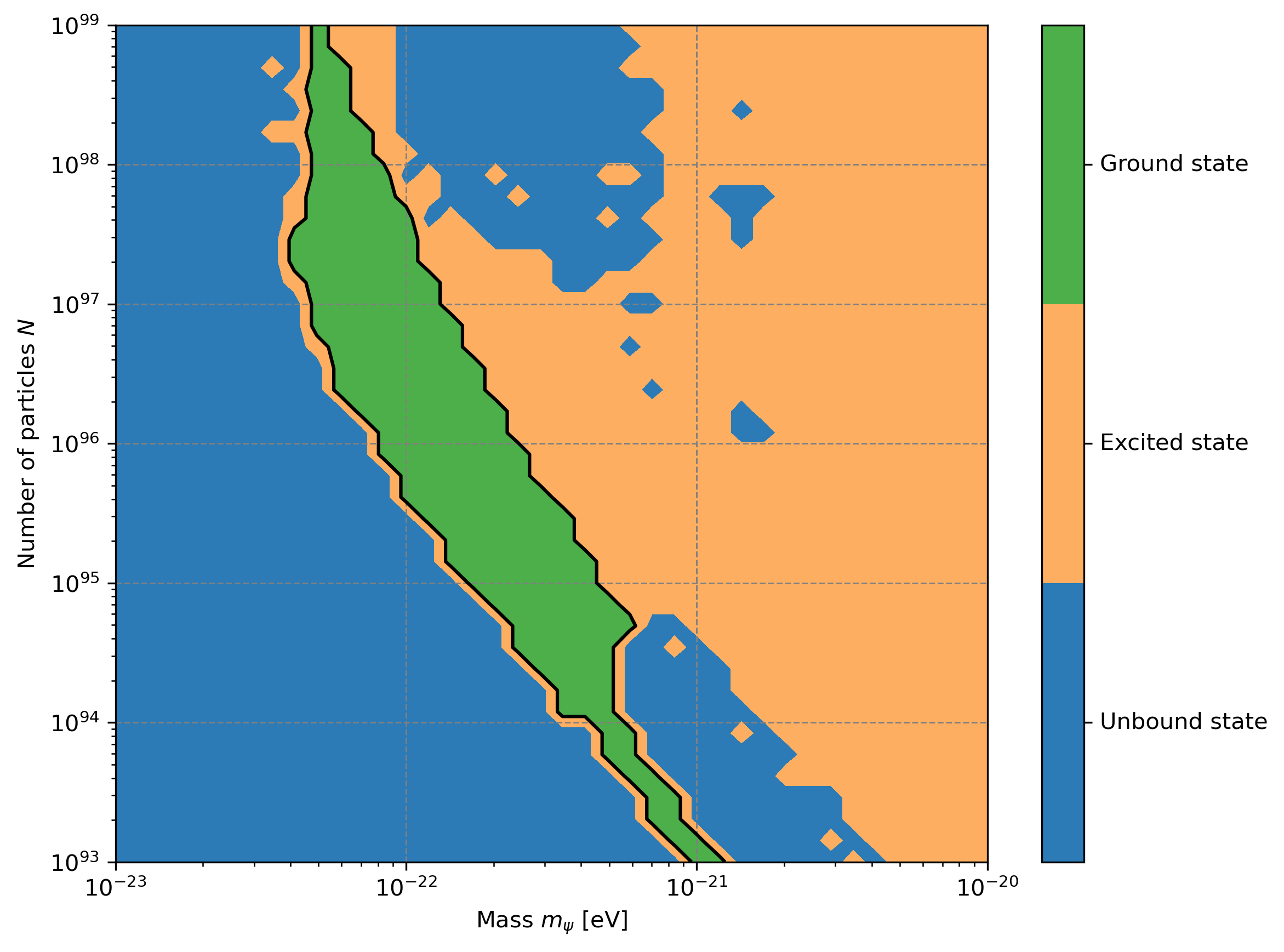}
    \caption{Gaussian ansatz}
    \label{gauss_77}
\end{subfigure}
\hfill
\begin{subfigure}{0.45\textwidth}
\centering
    \includegraphics[width=\linewidth]{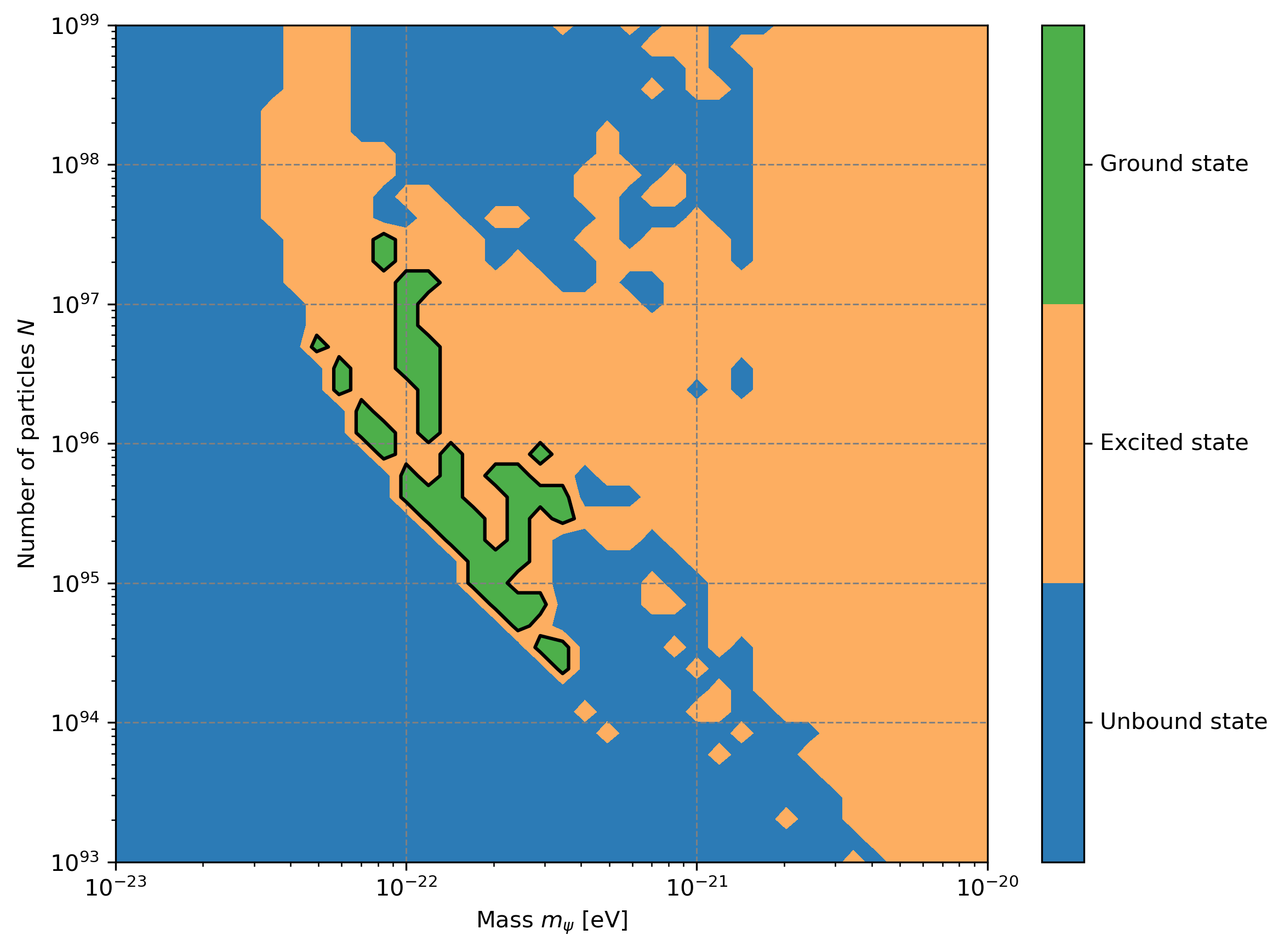}
    \caption{Exponential ansatz}
\end{subfigure}

\vspace{0.3cm}

\begin{subfigure}{0.45\textwidth}
\centering
    \includegraphics[width=\linewidth]{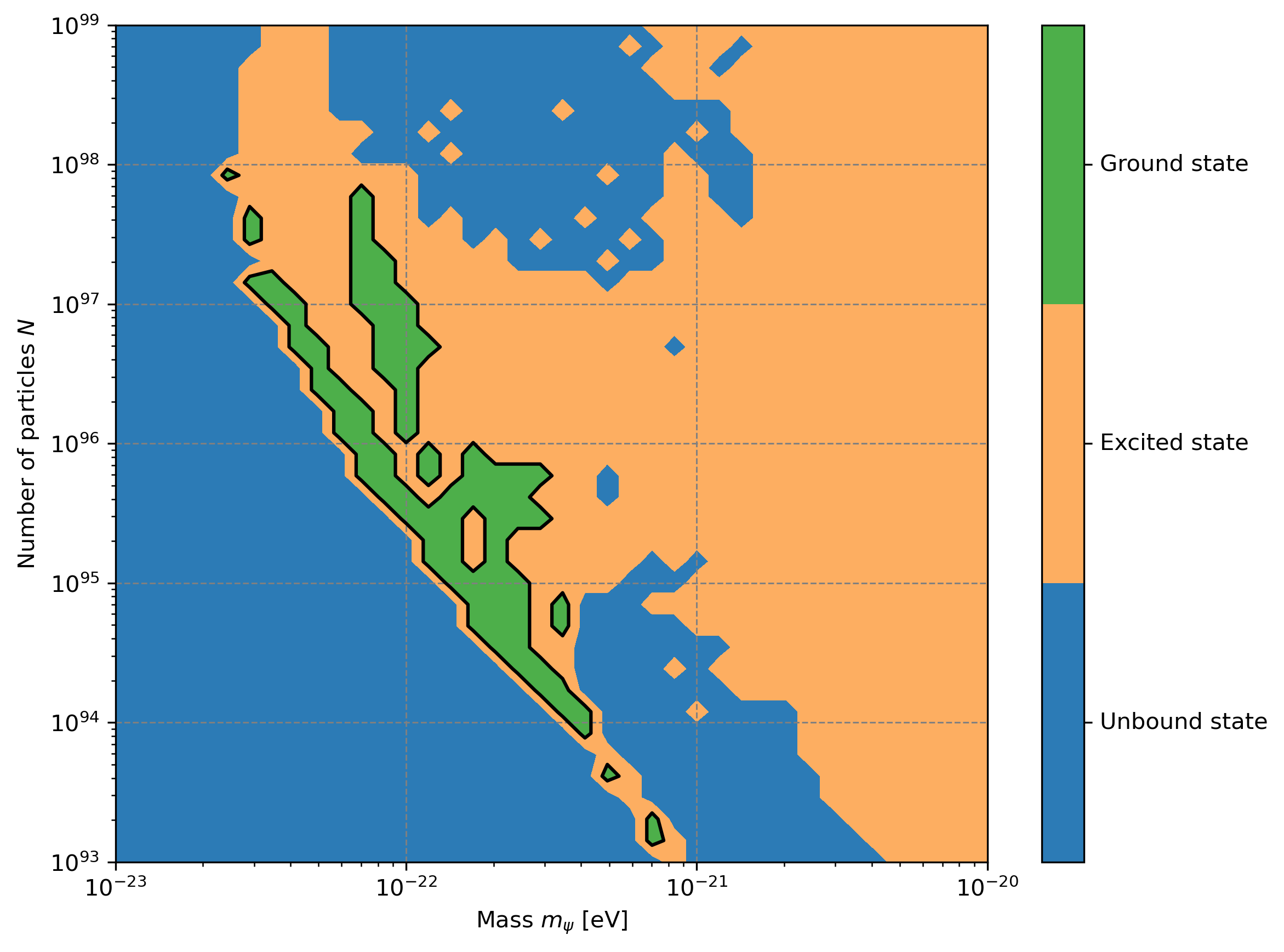}
    \caption{Linear–exponential ansatz}
\end{subfigure}
\hfill
\begin{subfigure}{0.45\textwidth}
\centering
    \includegraphics[width=\linewidth]{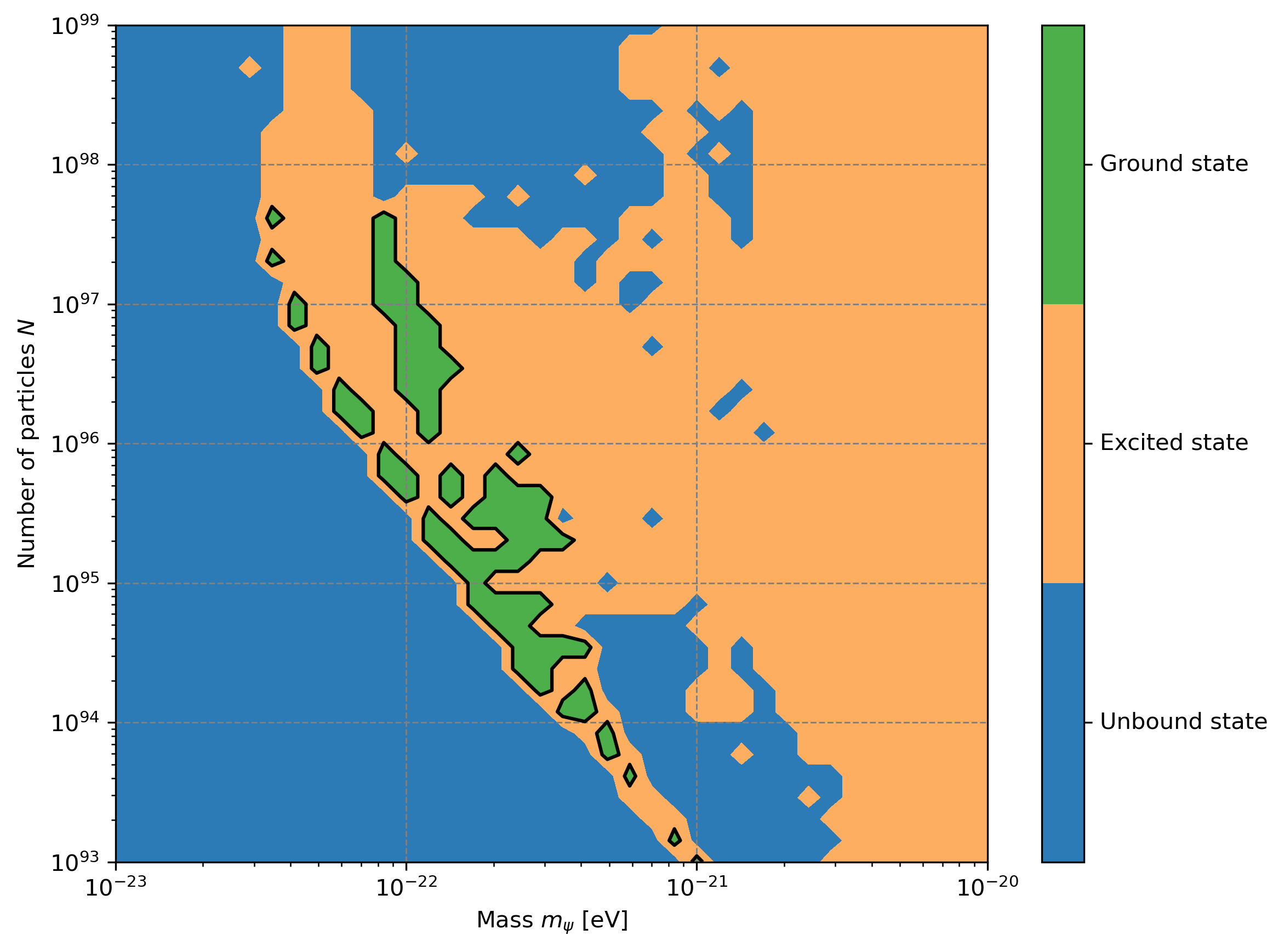}
    \caption{Hyperbolic secant ansatz}
\end{subfigure}
    \caption{
    Comparison of the $(m_{\phi},N)$ parameter space obtained for different ansätze of the initial wavefunction of the condensate state by using the Newton-Raphson method, with the scattering length fixed at $a_s = 10^{-77}\,\mathrm{m}$ and $r_d = 3\,\mathrm{kpc}$. Panels show results for the (a) Gaussian, (b) exponential, (c) linear–exponential, and (d) hyperbolic secant ansätze. Ground-state solutions are shown in green, excited-state solutions in orange, and unbound configurations in blue.
    }
\label{fig:ansatz_comparison}
\end{figure*}

\begin{figure*}[t]
\centering
\begin{subfigure}{0.48\textwidth}
\centering
    \includegraphics[width=\linewidth]{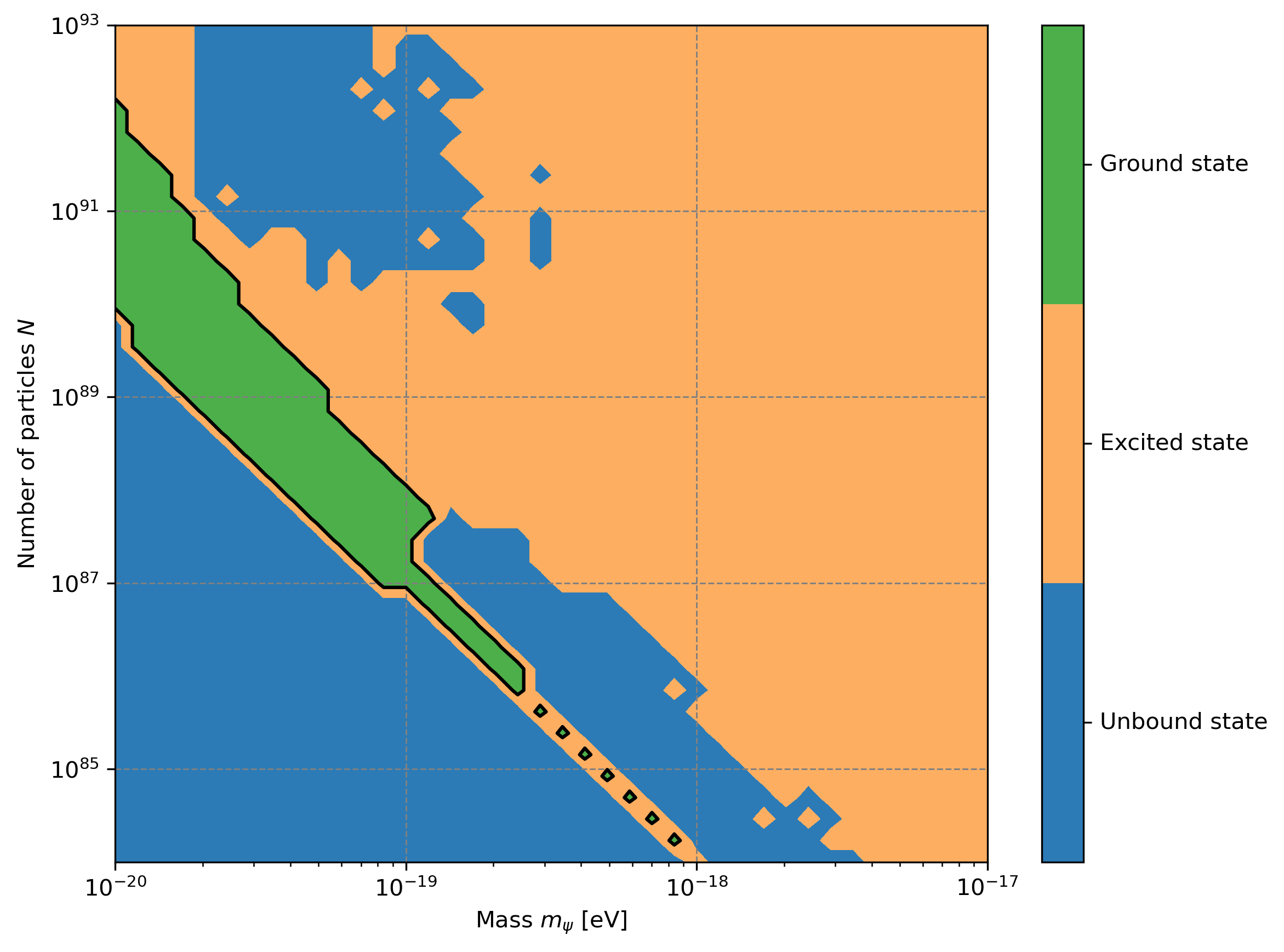}
    \caption{$a_s = 10^{-70}\,\mathrm{m}$.}
\label{gauss_70}
\end{subfigure}
\hfill
\begin{subfigure}{0.48\textwidth}
\centering
    \includegraphics[width=\linewidth]{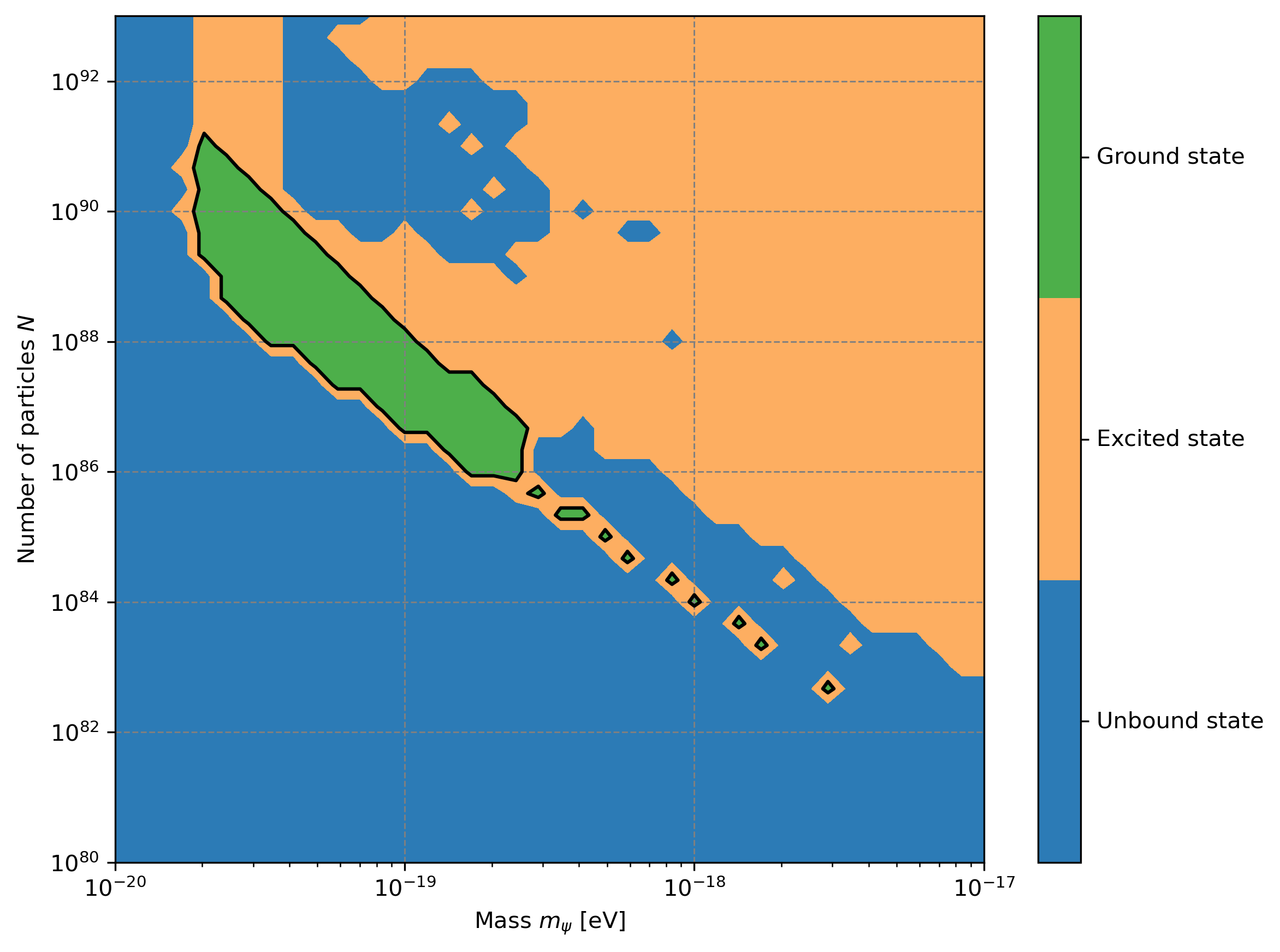}
    \caption{$a_s = 10^{-69}\,\mathrm{m}$.}
\label{gauss_69}
\end{subfigure}
\caption{
    Evolution of the $(m_{\phi},N)$ solution space obtained using a Gaussian ansatz as the scattering length is increased. Ground-state solutions are shown in green, excited-state solutions in orange, and unbound configurations in blue. The panels correspond to (a) $a_s = 10^{-70}\,\mathrm{m}$ and (b) $a_s = 10^{-69}\,\mathrm{m}$. Increasing the repulsive self-interaction shifts the ground-state branch toward larger boson masses and lower particle numbers.
} 
\label{fig:gauss_as_evolution}
\end{figure*}

\subsection{Structure of the $(m_\phi,N)$ solution space}\label{subsec:parameter_space}
The exploration of the $(m_{\phi},N)$ parameter space reveals the existence of three distinct classes of solutions: ground states, excited states, and unbound configurations. Figure~\ref{fig:ansatz_comparison} shows representative phase diagrams obtained using different initial ans\"atze while keeping the scattering length fixed at $a_s=10^{-77}\,\mathrm{m}$. We toke this value because ground-state solutions are around the fuzzy-dark-matter mass scale $m_\phi\sim10^{-22}\,\mathrm{eV}$, with particle numbers of order $N\sim10^{95}$--$10^{96}$.

A remarkable result is that the regions corresponding to excited states and unbound configurations are largely independent of the initial ansatz. This robustness indicates that these regions are intrinsic properties of the stationary GPP solution space rather than numerical artifacts associated with the numerical procedure.

As shown in Fig.~\ref{fig:ansatz_comparison}, the ground-state branch occupies an approximately diagonal region in the $(m_\phi,N)$ plane. Physically, this behavior reflects the competition between gravitational confinement and quantum support. Larger boson masses require fewer particles to generate a sufficiently deep gravitational potential well, whereas lighter bosons require significantly larger particle numbers in order to form bound equilibrium configurations.

The excited-state branch appears as an intermediate region separating ground-state and unbound configurations. These solutions satisfy the binding condition but belong to higher stationary branches of the nonlinear eigenvalue problem. Their presence is consistent with previous studies of the Schr\"odinger--Poisson and GPP systems, where stationary solutions with one or more radial nodes have been reported \cite{Guzman:2004wj,Guzman:2006yc,Schunck:2003kk}.

Configurations lying outside these regions are classified as unbound. In this regime the total energy per particle exceeds the depth of the self-generated gravitational potential well, preventing the formation of a localized self-gravitating condensate.

Although all four ans\"atze identify approximately the same excited-state and unbound regions, the Gaussian profile yields the largest continuous domain of convergence toward the ground-state branch. This suggests that the Newton--Raphson iteration is more robust when initialized with Gaussian profiles, making it the most efficient ansatz among those explored in this work.

Figures~\ref{gauss_77}, \ref{gauss_70}, and \ref{gauss_69} show that increasing the scattering length systematically displaces the ground-state branch toward larger boson masses and lower particle numbers. This behavior is expected because repulsive self-interactions provide an additional pressure-like contribution that partially counteracts gravitational collapse. Consequently, equilibrium configurations can be sustained with fewer particles than in the nearly non-interacting regime.

This trend can be understood from the balance between gravity, quantum pressure, and self-interactions. As $a_s$ increases, the repulsive interaction contributes increasingly to the support of the condensate, reducing the number of particles required to form a bound equilibrium configuration. The displacement of the ground-state branch therefore reflects a transition from configurations primarily supported by quantum pressure to configurations in which self-interactions play a significant role.

The comparison between Figs.~\ref{gauss_77}, \ref{gauss_70}, and \ref{gauss_69} also shows that the excited-state and unbound regions evolve more slowly with increasing $a_s$ than the ground-state branch, indicating that the location of the latter is considerably more sensitive to the strength of the self-interaction.

To our knowledge, previous studies of the GPP system have primarily characterized equilibrium solutions in terms of the halo mass and boson mass. The present analysis explicitly identifies the total particle number $N$ as an independent control parameter and determines empirical scaling relations linking $(m_\phi,N,a_s)$ to the observable halo radius $R_{99}$. These relations provide a practical bridge between the microscopic parameters of the condensate and macroscopic halo observables.

\begin{figure*}[tb]
\centering
    \includegraphics[width=0.48\textwidth]{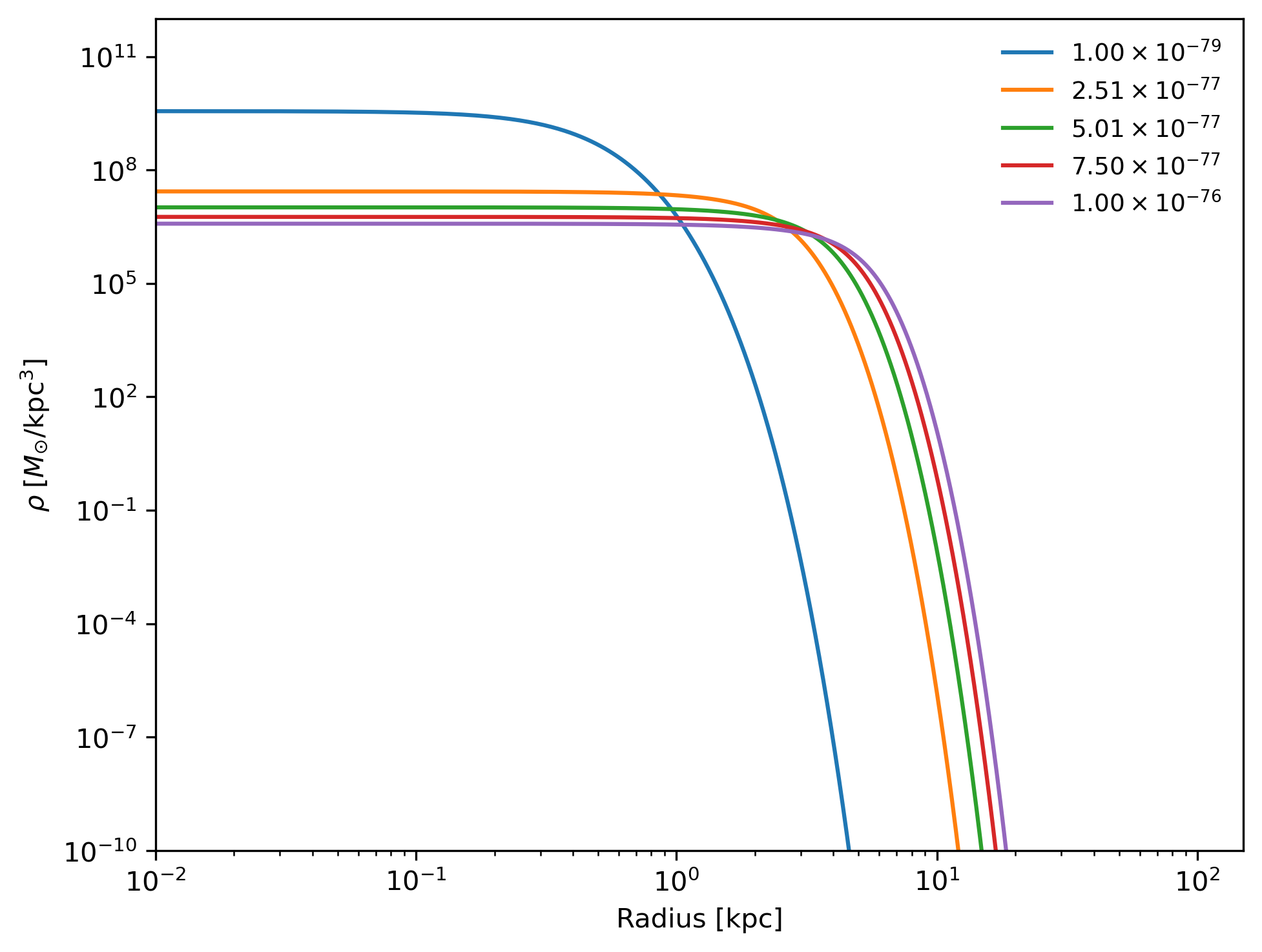}
    \hfill
    \includegraphics[width=0.48\textwidth]{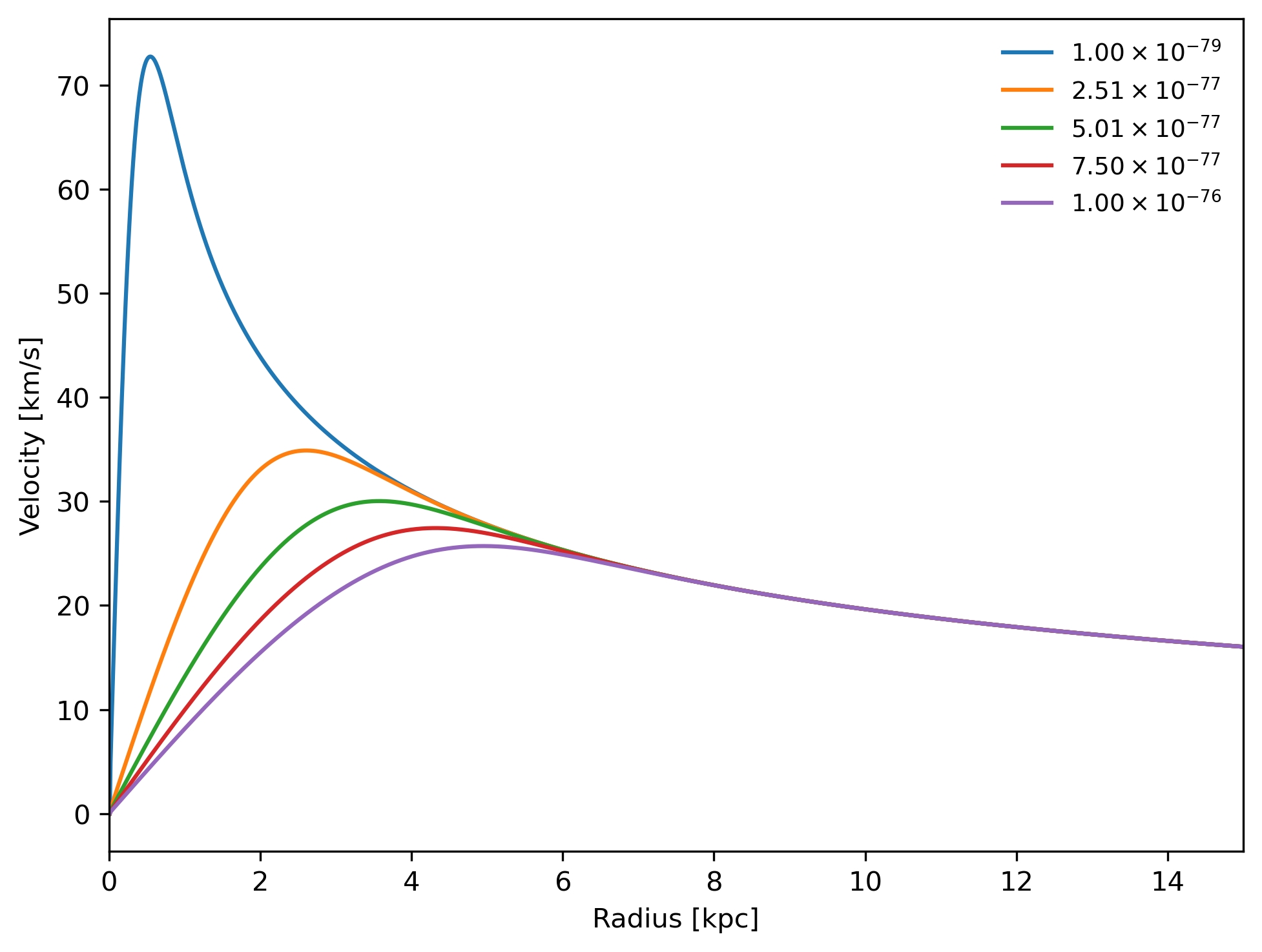}
\caption{
    \textit{Left:} Representative radial mass density profiles of ground-state halo solutions obtained with a Gaussian ansatz for different values of the scattering length $a_s$, shown in different colors. The boson mass $m_{\phi}$ and particle number $N$ are kept fixed. Larger values of $a_s$ correspond to more extended haloes. 
    \textit{Right:} Circular velocity profiles corresponding to the density distributions. Increasing the scattering length $a_s$ results in lower peak velocities and a broader radial extension of the halo.
}
\label{fig:gauss_dif_a}
\end{figure*}

\begin{table}[t]
    \centering    
    \begin{tabular}{cc}
        \toprule
        $a$ [m] & $R_{99}$ [kpc] \\
        \midrule
        $1\times10^{-79}$  & 1.01 \\
        $2.5\times10^{-77}$  & 3.83\\
        $5\times10^{-77}$  & 5.05 \\
        $7.5\times10^{-77}$  & 5.98 \\
        $1\times10^{-76}$ & 6.76 \\
        \bottomrule
    \end{tabular}
    \caption{Radius $R_{99}$ enclosing $99\%$ of the total halo mass for different values of the scattering length $a_s$, obtained from ground-state solutions with fixed $m_{\phi} = 10^{-22}\,\mathrm{eV}$ and $N = 10^{97}$, corresponding to the solutions shown in Fig.~\ref{fig:gauss_dif_a}. The values of $R_{99}$ increase with increasing $a_s$.} 
    \label{tab:R99_dif_a}
\end{table}
%
\subsection{Ground-state halo profiles}
To understand how the particle properties and the total particle number affect the structure of self-gravitating condensates, we first examine representative ground-state solutions obtained by varying the scattering length, the particle number, and the boson mass. These examples provide a direct visualization of how the density distribution, circular velocity profile, and characteristic halo size respond to changes in the physical parameters of the GPP system.

Figure~\ref{fig:gauss_dif_a} shows representative ground-state density and circular-velocity profiles obtained for different values of the scattering length while keeping $m_\phi$ and $N$ fixed. The corresponding values of the characteristic radius $R_{99}$ are listed in Table~\ref{tab:R99_dif_a}.

As the scattering length increases, the condensate becomes more extended and less centrally concentrated. Consequently, the characteristic halo size increases from approximately $1\,\mathrm{kpc}$ to nearly $7\,\mathrm{kpc}$ across the interval explored. At the same time, the peak circular velocity decreases because the mass is distributed over a larger volume. In general, the density profiles exhibit an approximately constant-density central region before decreasing at larger radii.

Figure~\ref{fig:gauss_dif_N} illustrates the effect of varying the total particle number while keeping $m_\phi$ and $a_s$ fixed. The corresponding halo properties are summarized in Table~\ref{tab:R99_dif_N}.

Increasing the particle number increases the total halo mass according to $M_H=m_\phi N$. As a consequence, the gravitational potential becomes deeper and the equilibrium configuration contracts. This behavior is reflected by the decrease of $R_{99}$ and the increase of the peak circular velocity as $N$ grows.

Figure~\ref{fig:gauss_dif_m} shows the dependence of the ground-state solutions on the boson mass, while Table~\ref{tab:R99_dif_m} summarizes the corresponding values of $R_{99}$ and $M_H$.

Increasing the boson mass produces substantially more compact configurations. The characteristic radius decreases by more than one order of magnitude across the explored mass range, while the central density and peak circular velocity increase. This behavior reflects the stronger gravitational confinement associated with larger boson masses.

\subsection{Scaling relations of ground-state halo configurations}
One of the main outcomes of this work is the determination of empirical scaling relations connecting the particle properties $(m_{\phi},a_s)$ and the macroscopic control parameter $N$ to the halo properties of the resulting halo-like configurations. Using the set of converged ground-state solutions, we performed ordinary least-squares fits to determine how the characteristic halo radius $R_{99}$ depends on the physical parameters of the model.

\begin{figure*}[t]
\centering
    \includegraphics[width=0.48\textwidth]{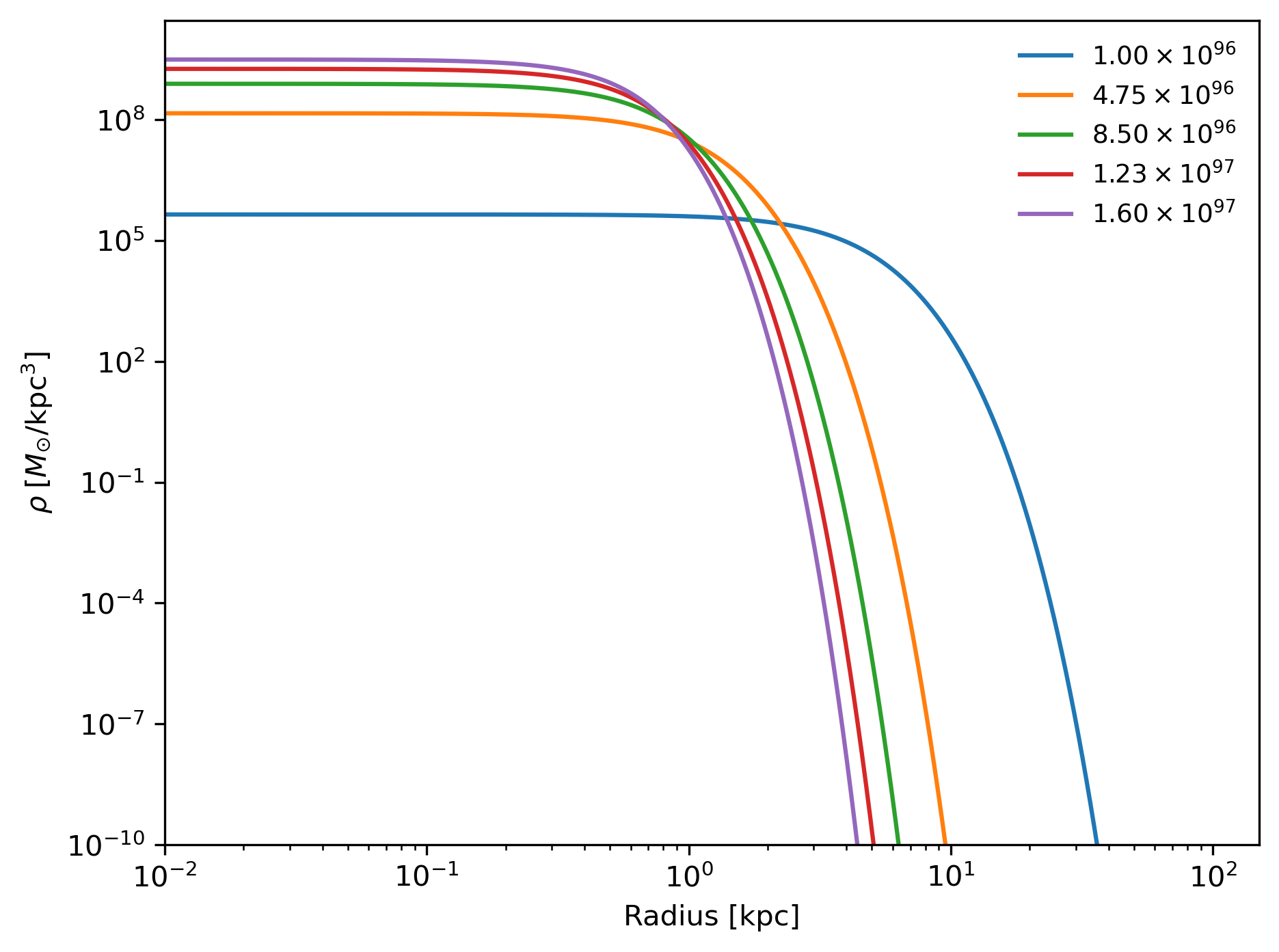}
    \hfill
    \includegraphics[width=0.48\textwidth]{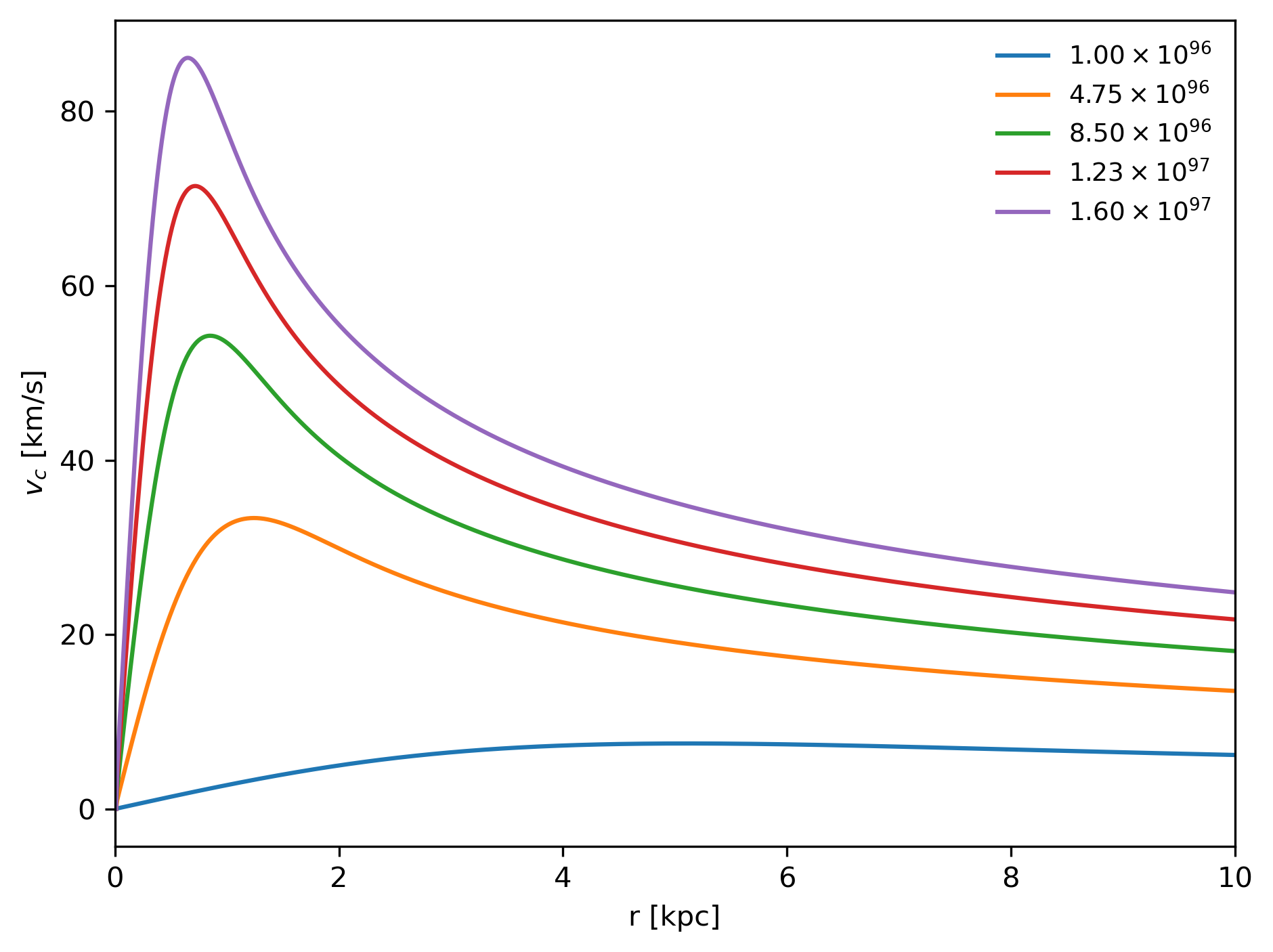}
\caption{
    \textit{Left:} Representative radial mass density profiles of ground-state halo solutions obtained with a Gaussian ansatz for different values of the particle number $N$, keeping the boson mass fixed at $m_{\phi}=10^{-22}\,\mathrm{eV}$ and the scattering length at $a_{s}=10^{-78}\,\mathrm{m}$. 
    \textit{Right:} Circular velocity profiles corresponding to the density distributions. Increasing the particle number produces more massive haloes and therefore higher circular velocities.
}
\label{fig:gauss_dif_N}
\end{figure*}

\begin{table}[t]
    \centering
        \begin{tabular}{ccc}
        \toprule
        $N$ & $R_{99}$ [kpc] & $M_{\textrm{H}}$ [$M_{\odot}$] \\
        \midrule
        $1\times10^{96}$  & 9.55 & $8.97\times10^{7}$\\
        $4.75\times10^{96}$  & 2.27 & $4.26\times10^{8}$ \\
        $8.50\times10^{96}$  & 1.50& $7.62\times10^{8}$ \\
        $1.23\times10^{97}$  & 1.23 & $1.10\times10^{9}$ \\
        $1.6\times10^{97}$ & 1.09 & $1.43\times10^{9}$\\
        \bottomrule
    \end{tabular}
    \caption{$R_{99}$ for different values of $N$, obtained from ground-state solutions with fixed $m_{\phi}$ and $a_s$, corresponding to the solutions shown in Fig~\ref{fig:gauss_dif_N}. $R_{99}$ decrease with increasing the particle number $N$.} 
    \label{tab:R99_dif_N}
\end{table}

In the region corresponding to $10^{-23}\,\mathrm{eV} \leq m_{\phi} \leq 10^{-22}\,\mathrm{eV}$, $10^{96} \leq N \leq 10^{97}$, 
with $a_s=10^{-80}\,\mathrm{m}$, we obtain
\begin{equation}
    R_{99} = R_{\star} \left( \frac{m_{\phi}} {10^{-22}\,\mathrm{eV}} \right)^{-2.996\pm0.001} \left( \frac{N} {10^{97}} \right)^{-0.998\pm0.001},
\end{equation}
where
$R_{\star} = 0.953\pm0.001~\mathrm{kpc}$. The fitted exponents are remarkably close to $R\propto m_{\phi}^{-3}N^{-1}$.  Since the total halo mass satisfies $M_{H}=m_{\phi}N$, this relation can be rewritten as
\begin{equation}
    R_{99}\propto \frac{1}{m_{\phi}^{2}M_{H}},
\end{equation}
which coincides with the well-known mass--radius relation of non-interacting Schr\"odinger--Poisson solitons. This relation can be derived analytically from the virial equilibrium and assuming a negligible self-interaction contribution \cite{Chavanis:2011zi,Hui:2016ltb}. This agreement provides a consistency check of the numerical implementation.

When the repulsive self-interactions are included in the fitting, the scattering length becomes an additional scale controlling the equilibrium radius. Using the same mass region $10^{-23}\,\mathrm{eV} \leq m_{\phi} \leq 10^{-22}\,\mathrm{eV}$ and $10^{-78} \textrm{ m} \leq a_{s} \leq 10^{-77} \textrm{ m}$ with $N \sim 10^{100}$ we obtain
\begin{equation}
    R_{99} = R_{\circ} \left( \frac{a_{s}} {10^{-77} \textrm{ m}} \right)^{0.481 \pm 0.003} \left( \frac{m_{\phi}} {10^{-22}\,\mathrm{eV}} \right)^{-1.545 \pm 0.008} ,
\end{equation}
with $R_{\circ} = 1.583 \pm 0.014 \textrm{ kpc}$. The scaling obtained at fixed $N$, $R_{99}\propto a_s^{0.481}m_{\phi}^{-1.545}$,
resembles the Thomas--Fermi expectation
\begin{equation}
    R_{\rm TF}\propto \left(\frac{a_s}{m_{\phi}^3}\right)^{1/2} = a_s^{1/2}m_{\phi}^{-3/2},
\end{equation}
showing that repulsive self-interactions play an important role in setting the halo size. The fitted exponents are remarkably close to the Thomas-Fermi prediction, when the kinectic energy contribution is negligible, although small deviations remain detectable. These deviations suggest that quantum-pressure effects still contribute to the equilibrium structure \cite{Boehmer:2007um,Chavanis:2011zi,Chavanis:2011uv}.

Allowing all three parameters to vary simultaneously in the region $2\times10^{-23} \textrm{ eV} \leq m_{\phi} \leq 2\times10^{-22} \textrm{ eV}$, $10^{96} \leq N \leq 10^{97}$ and $10^{-78} \textrm{ m} \leq a_{s} \leq 10^{-77} \textrm{ m}$ we get the fit
\begin{equation}
\scriptsize
    R_{99} = R_{\triangle}\left( \frac{m_{\phi}} {10^{-22}\,\mathrm{eV}} \right)^{-2.096 \pm 0.008} \left( \frac{a_{s}} {10^{-77}\,\mathrm{m}} \right)^{0.308 \pm 0.004} \left( \frac{N} {10^{97}} \right)^{-0.391 \pm 0.004}.
\end{equation}
with $R_{\triangle} = 2.848 \pm 0.019\textrm{ kpc}$. This reveals that the radius is most sensitive to the boson mass, moderately sensitive to the particle number and softly dependace to the scattering length.

\begin{figure*}[t]
\centering
    \includegraphics[width=0.48\textwidth]{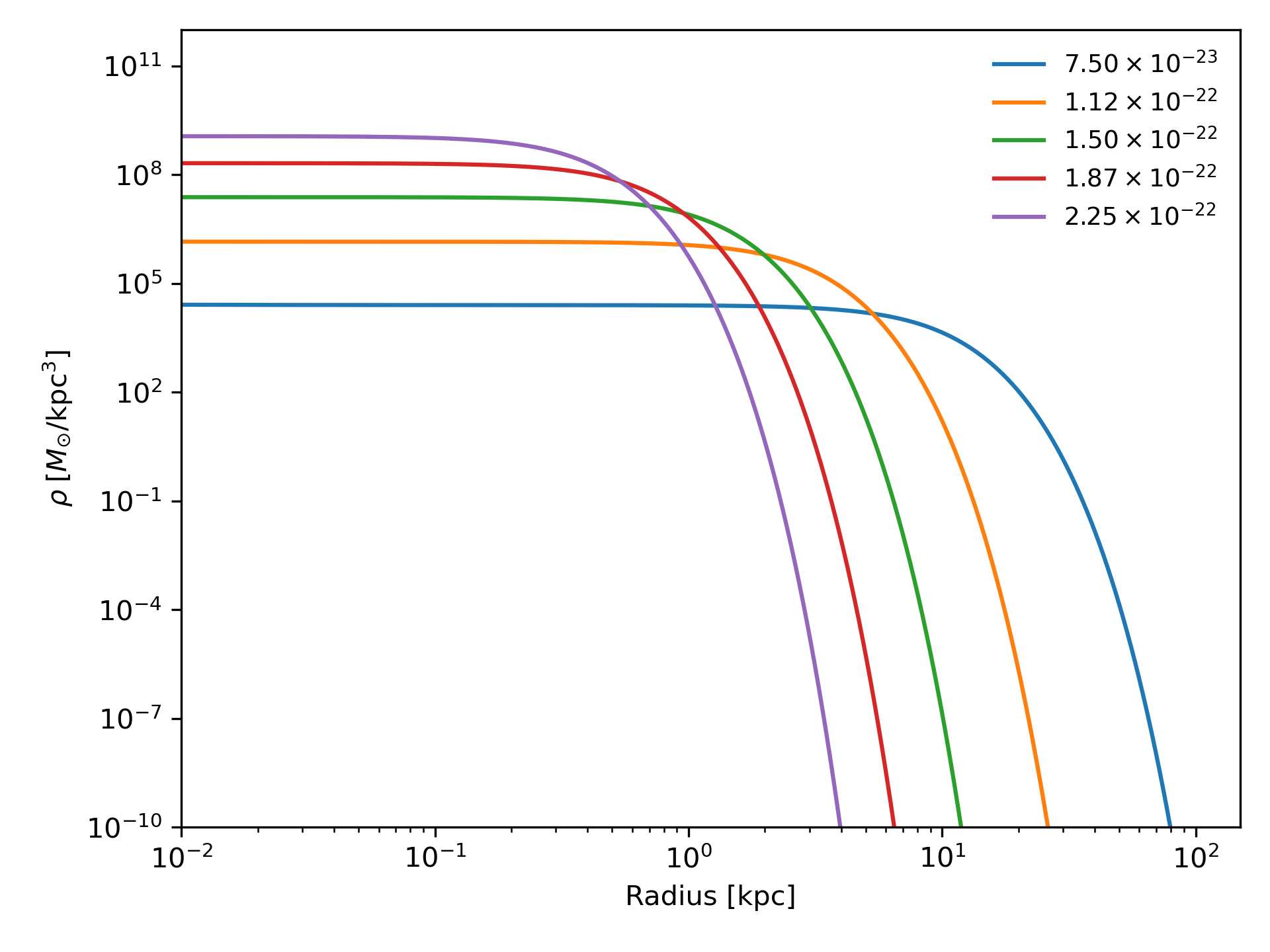}
    \hfill
    \includegraphics[width=0.48\textwidth]{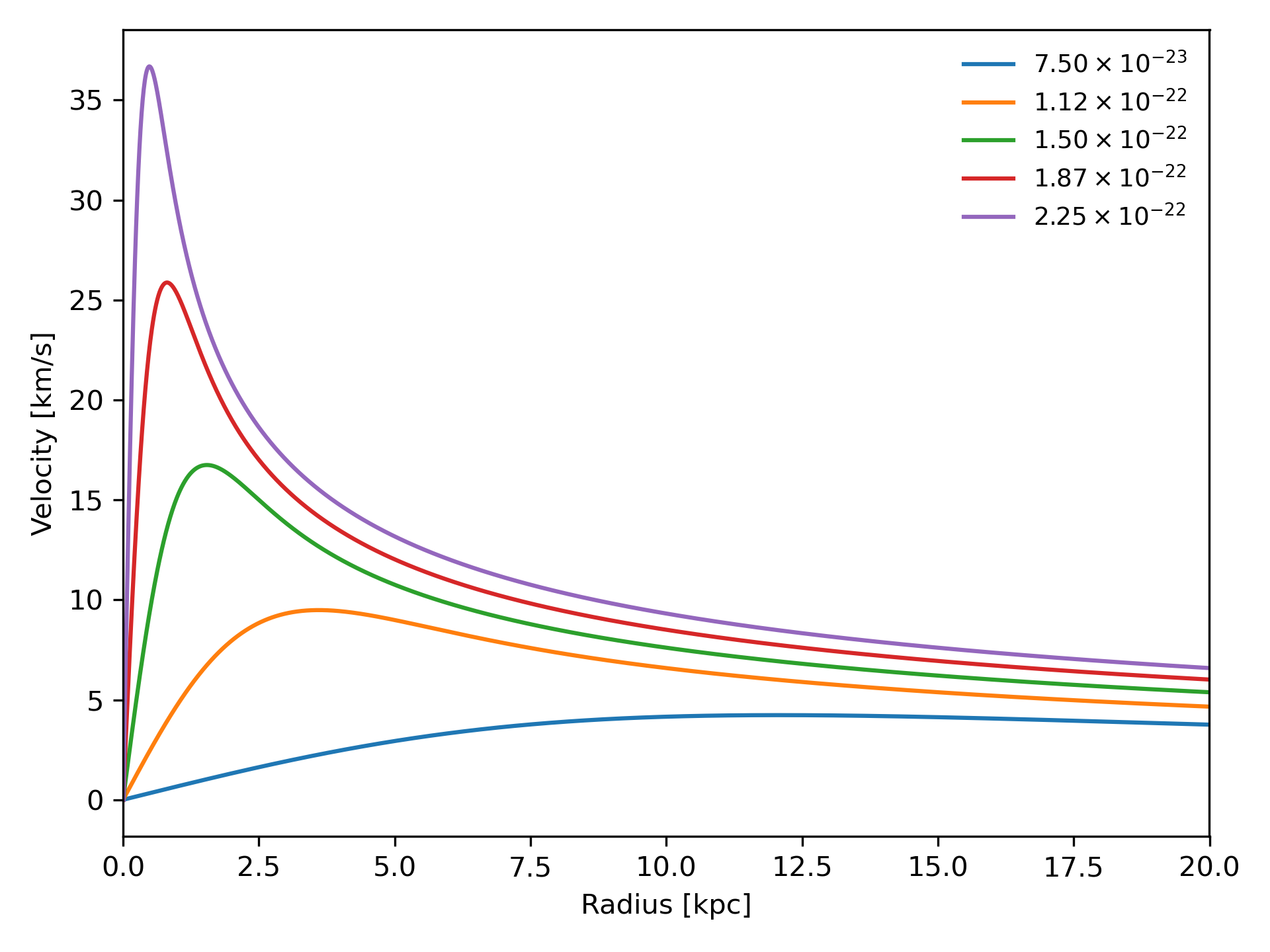}
    \caption{
    \textit{Left:} Representative radial mass density profiles of ground-state halo solutions obtained with a Gaussian ansatz for different values of the boson mass $m$, shown in different colors. The particle number $N=10^{96}$ and the scattering length $a_s=10^{-78}\,\mathrm{m}$ are kept fixed. 
    \textit{Right:} Circular velocity profiles corresponding to the density distributions. Increasing the boson mass $m$ leads to more compact haloes and therefore higher peak circular velocities.
}
\label{fig:gauss_dif_m}
\end{figure*}

These results quantitatively confirm the trends already observed in Figs.\ref{fig:gauss_dif_a}--\ref{fig:gauss_dif_m}: increasing the scattering length produces larger and more diffuse halos, whereas increasing either the boson mass or the particle number leads to more compact configurations.

\begin{table}[t]
    \centering    
    \begin{tabular}{ccc}
        \toprule
        $m_{\phi}$[eV] & $R_{99}$ [kpc] & $M_{\textrm{H}}$ [$M_{\odot}$] \\
        \midrule
        $7.50\times10^{-23}$  & 22.54 & $6.72\times10^{7}$\\
        $1.12\times10^{-22}$  & 6.73 & $1.01\times10^{8}$ \\
        $1.5\times10^{-22}$  & 2.88& $1.34\times10^{8}$ \\
        $1.87\times10^{-22}$  & 1.51 & $1.68\times10^{8}$ \\
        $2.25\times10^{-22}$ & 0.90 & $2.02\times10^{8}$\\
        \bottomrule
    \end{tabular}
    \caption{$R_{99}$ for different values of $m_{\phi}$, obtained from ground-state solutions with fixed $N$ and $a_s$, corresponding to the solutions shown in Fig~\ref{fig:gauss_dif_m}. $R_{99}$ decrease with increasing the mass of the boson $m_{\phi}$.}
    \label{tab:R99_dif_m}
\end{table}

Taken together, the fitted relations reveal a clear hierarchy among the microscopic parameters. The boson mass has the strongest impact on the halo size, followed by the total particle number, while the scattering length acts as a secondary but non-negligible regulator of the equilibrium radius. These results quantify how quantum pressure, self-interactions, and gravity jointly determine the structure of stationary GPP halos.

\subsection{Excited-state configurations}
\begin{figure}[t]
    \centering
    \includegraphics[width=0.5\textwidth]{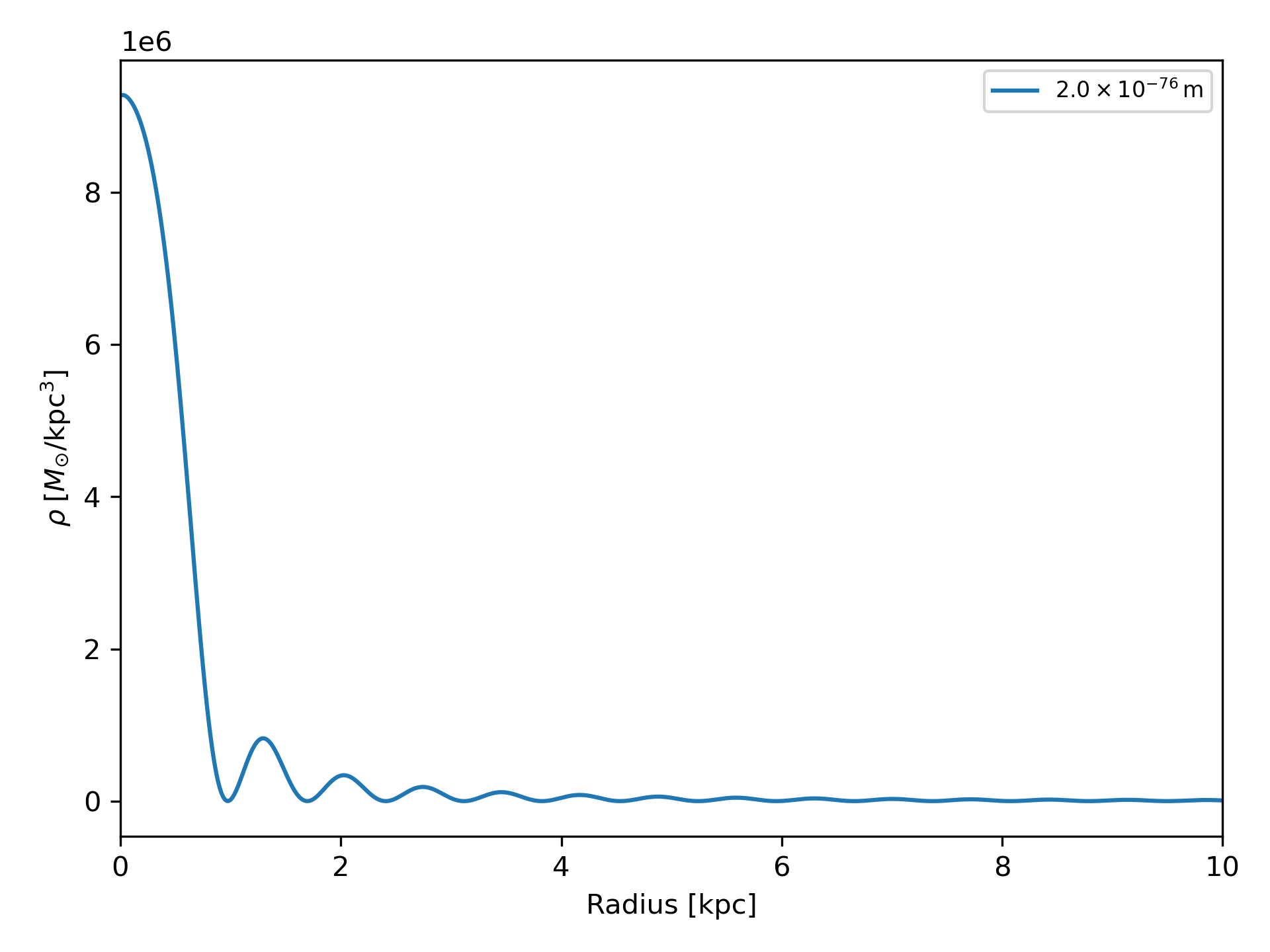} 
        \caption{Radial density profile of a representative excited-state solution for $m_{\phi}=10^{-22}\,\mathrm{eV}$, $N=10^{97}$, and $a_s=2\times10^{-76}\,\mathrm{m}$. The oscillatory structure reflects the nodal character of the underlying condensate wavefunction.}
    \label{fig:gauss_rho_w_nodes}
\end{figure}

In addition to ground-state solutions, the numerical exploration of the GPP system reveals the existence of excited-state configurations. Figure~\ref{fig:gauss_rho_w_nodes} shows a representative excited-state solution obtained for $m_\phi=10^{-22}\,\mathrm{eV}$, $N=10^{97}$, and $a_s=2\times10^{-76}\,\mathrm{m}$.

Unlike ground-state solutions, whose order parameter remains positive throughout the computational domain, excited states are characterized by one or more radial nodes where the condensate wavefunction vanishes and changes sign. These nodes arise naturally from the nonlinear eigenvalue structure of the stationary GPP equations and are analogous to the excited eigenstates of the linear Schr\"odinger equation. Although the mass density $\rho=m_{\phi}|\psi|^2$ remains positive everywhere, the nodal structure of $\psi$ identifies the solution as belonging to a higher-energy stationary branch.

The corresponding density profile exhibits a central core surrounded by a sequence of oscillatory density shells whose amplitude decreases with radius. Similar excited-state solutions have been reported previously in studies of the Schr\"odinger--Poisson and GPP systems, where they appear as stationary configurations with one or more radial nodes \cite{Guzman:2004wj,Guzman:2006yc,Schunck:2003kk,Chavanis:2025qcg}. In the boson-star literature these solutions are commonly interpreted as excited equilibrium states of the condensate.

From a dynamical perspective, excited states are generally expected to be less stable than the ground-state branch. Time-dependent simulations have shown that excited configurations tend to lose mass through gravitational cooling and progressively migrate toward lower-energy states, ultimately approaching the ground-state soliton \cite{Guzman:2004wj,Guzman:2006yc}. For this reason, the ground-state branch is usually regarded as the physically preferred equilibrium configuration.

In the present work, excited states are retained as valid stationary solutions of the nonlinear eigenvalue problem and constitute an important component of the solution space. Their existence delineates the transition between ground-state and unbound configurations in the $(m_{\phi},N,a_s)$ parameter space and provides additional insight into the structure of the stationary GPP spectrum.

\subsection{Rotation curves}
As an astrophysical application of the stationary solutions obtained in this work, we investigate whether ground-state GPP configurations can reproduce observed galactic rotation curves. For this purpose we use galaxies from the SPARC catalog \cite{Lelli_2016}, whose high-quality rotation curves provide a stringent test of dark-matter halo models.

The circular velocity generated by the condensate halo is computed from the enclosed mass profile according to according to Eq.~(\ref{eq:vc_2}) in Appendix~\ref{app:rotation_curves}. The total rotation curve is modeled as the sum of the baryonic contribution (gas, stellar disk, and bulge when present) and the dark-matter contribution provided by the stationary GPP solution. Throughout this analysis we adopt a stellar mass-to-light ratio $\Upsilon_{\star}=0.574$.

\begin{figure*}[t]
\centering
\begin{subfigure}{0.45\textwidth}
\centering
    \includegraphics[width=1\textwidth]{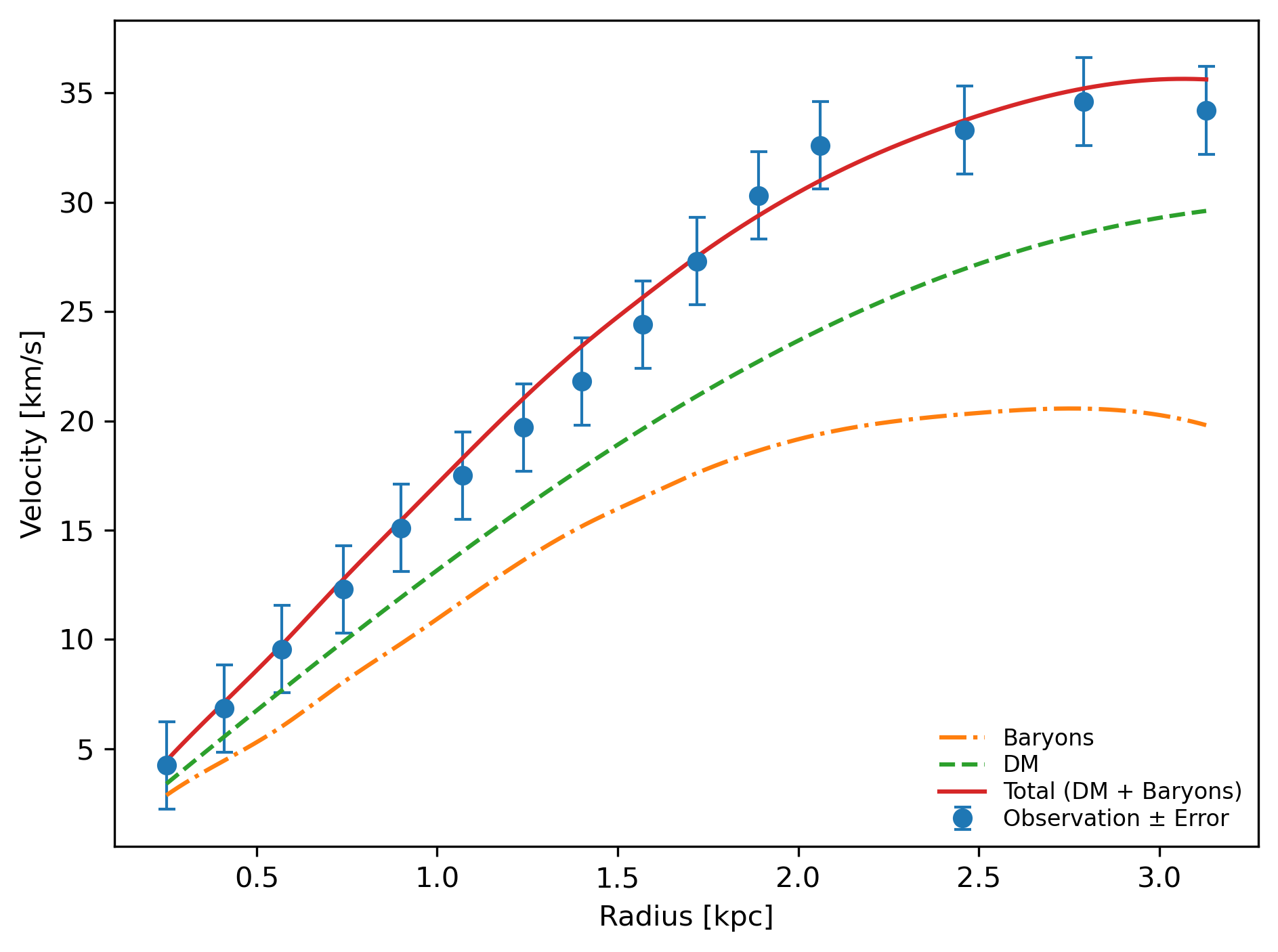}
    \caption{KK98-251}
\label{KK98_251_G}
\end{subfigure}
\hfill
\begin{subfigure}{0.45\textwidth}
\centering
    \includegraphics[width=1\textwidth]{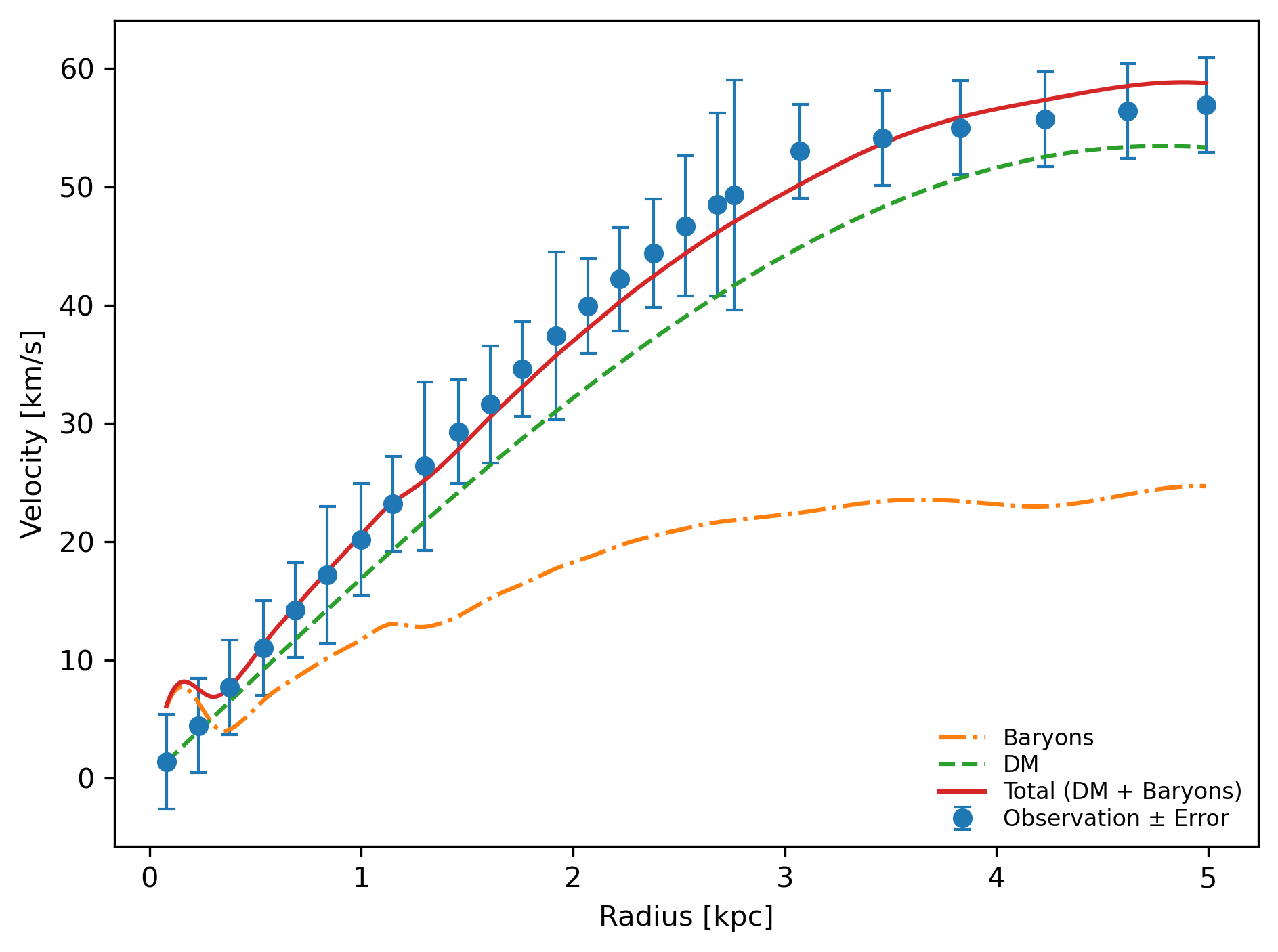}
    \caption{UGC01281}
\label{UGC01281_G}
\end{subfigure}
\caption{
    Ground-state solutions of the GPP system used as the dark matter component of the rotation curves, obtained with a Gaussian ansatz. The blue points and bars are the observed data with their respective error, the dash dot orange line correspond to the baryonic contribution; gas, stars and bulge (if any), while dash green line is the soliton as the DM halo and continuous red line is the total rotation curve of the galaxy.
    \textit{(a)} Galaxy KK98-251. The corresponding soliton halo has a total mass of $8.96\times10^{8}M_{\odot}$ and $R_{99}=5.05\,\mathrm{kpc}$ for $m_{\phi}=10^{-22}\,\mathrm{eV}$, $N=10^{97}$ and $a_s=5\times10^{-77}\,\mathrm{m}$. 
    \textit{(b)} Galaxy UGC01281. The soliton halo has a mass of $3.58\times10^{9}M_{\odot}$ and $R_{99}=5.88\,\mathrm{kpc}$ for $m_{\phi}=10^{-22}\,\mathrm{eV}$, $N=4\times10^{97}$ and $a_s=10^{-76}\,\mathrm{m}$. 
}
\label{fig:gauss_galaxies}
\end{figure*}

Figure~\ref{fig:gauss_galaxies} presents two representative examples, KK98-251 and UGC01281, obtained using ground-state solutions generated from a Gaussian ansatz with a fixed boson mass $m_{\phi}=10^{-22}\,\mathrm{eV}$. The corresponding halo parameters are listed in Table~\ref{tab:galaxies}.

\begin{table}[t]
\centering    
    \begin{tabular}{lcccc}
    \toprule
    Galaxy & $N$ & $a_s$ & $R_{99}$ & $M_H$ \\
    &  & [m] & [kpc] & [$10^9M_\odot$] \\
    \midrule
    KK98-251 & $1\times10^{97}$ & $5\times10^{-77}$ & 5.05 & 0.896 \\
    UGC01281 & $4\times10^{97}$ & $1\times10^{-76}$ & 5.88 & 3.58 \\
    \bottomrule
    \end{tabular}
\caption{
    Parameters of the ground-state GPP halos used to reproduce the rotation curves shown in Fig.~\ref{fig:gauss_galaxies}. The boson mass is fixed to $m_\phi=10^{-22}\,\mathrm{eV}$.
    }
\label{tab:galaxies}
\end{table}

For KK98-251, the best-fit configuration corresponds to a halo mass $M_{\mathrm H}=8.96\times10^8\,M_{\odot}$ and a characteristic radius $R_{99}=5.05\,\mathrm{kpc}$. For UGC01281, the resulting halo mass is $M_{\mathrm H}=3.58\times10^9\,M_{\odot}$ with $R_{99}=5.88\,\mathrm{kpc}$. In both cases the resulting rotation curves provide a satisfactory description of the observed kinematics over the radial range probed by the data.

An interesting feature of these solutions is that the observed rotation curves can be reproduced using only the ground-state condensate configuration. In contrast to the standard picture emerging from cosmological simulations of fuzzy dark matter, where halos are composed of a central solitonic core embedded within an extended envelope \cite{Schive:2014hza,Hui:2016ltb}, the galaxies considered here can be described by the solitonic component alone. No additional halo envelope is required to reproduce the available rotation-curve measurements.

The characteristic halo sizes inferred from the GPP solutions are also compatible with the typical scales reported in observational studies of dwarf galaxies. For example, the value $R_{99}=5.05\,\mathrm{kpc}$ obtained for KK98-251 is comparable to the core radii inferred from pseudo-isothermal fits reported in Refs.~\cite{Begum_2004,Lelli_2016}, although the quantities are not strictly equivalent.

These results demonstrate that stationary ground-state solutions of the GPP system can generate realistic galactic rotation curves while maintaining physically plausible halo masses and sizes. They therefore provide a direct connection between the particle properties, the total particle content, and observable galactic dynamics.

\subsubsection{Implications for Lyman-$\alpha$ constraints}

One of the main challenges faced by fuzzy and ultralight dark matter models arises from observations of the Lyman-$\alpha$ forest. Hydrodynamical analyses of the Lyman-$\alpha$ flux power spectrum have placed increasingly stringent lower bounds on the boson mass, typically excluding the canonical fuzzy-dark-matter value $m_{\phi}\sim10^{-22}\,\mathrm{eV}$ when the ultralight component constitutes the dominant dark matter fraction \cite{Irsic:2017yje,Armengaud:2017nkf,Rogers:2020ltq}. Depending on the dataset and modeling assumptions, these studies generally require $m_{\phi}\gtrsim 10^{-21}\text{--}10^{-20}\,\mathrm{eV}$, placing the traditional fuzzy-dark-matter scenario in tension with small-scale structure constraints.

As discussed in Sec.~\ref{subsec:parameter_space} and shown in Fig.~\ref{fig:gauss_as_evolution}, increasing the scattering length shifts the ground-state branch toward larger boson masses and smaller particle numbers. This behavior allows stationary GPP solutions to exist in regions of parameter space compatible with current Lyman-\(\alpha\) constraints.

For scattering lengths of order $a_s \sim 10^{-69}\,\mathrm{m}$, the ground-state branch extends into the mass range
\begin{equation}
    m_\phi\simeq2.03\times10^{-20}-2.89\times10^{-18}\,\mathrm{eV},
\end{equation}
which is compatible with current Lyman-$\alpha$ constraints. The corresponding solutions are obtained for particle numbers in the approximate range $N\simeq4.64\times10^{82}-10^{91}$.

Although these results demonstrate that the stationary GPP system admits self-gravitating equilibrium solutions in the mass range favored by Lyman-$\alpha$ observations, the resulting halo configurations differ substantially from those required to reproduce the dwarf-galaxy rotation curves analyzed in Sec.\ref{sec:results}. In particular, the circular velocities generated by these high-mass solutions are significantly smaller than those observed in galaxies such as KK98-251 and UGC01281.

Therefore, within the class of stationary halo configurations considered here, the tension between the boson masses preferred by dwarf-galaxy rotation curves $m_{\phi}\sim10^{-22}\,\mathrm{eV}$ and those favored by Lyman-$\alpha$ forest observations $m_{\phi}\gtrsim10^{-20},\mathrm{eV}$ remains unresolved. Nevertheless, the shift of the ground-state region toward larger masses as the scattering length increases highlights the important role of self-interactions in determining the phenomenology of self-gravitating Bose--Einstein-condensate dark matter halos.

\section{Conclusions}\label{sec:conclusions}

In this work we have constructed and analyzed stationary halo-like solutions of the Gross--Pitaevskii--Poisson system for self-gravitating bosons with weak repulsive self-interactions. The stationary equations were solved in spherical symmetry using a finite-difference discretization and a Newton--Raphson iterative scheme, while keeping the particle properties $(m_\phi,a_s)$ and the total particle number $N$ explicit throughout the calculation. This formulation allows a direct connection between the particle mass, the total number of bosons, the scattering length, and the macroscopic properties of the resulting equilibrium configurations.

A central result of our analysis is the classification of the stationary solution space into ground states, excited states, and unbound configurations. By scanning the $(m_\phi,N)$ parameter space for fixed values of the scattering length, we found that the ground-state branch occupies a well-defined region whose location depends sensitively on $a_s$. Increasing the scattering length shifts this region toward larger boson masses and smaller particle numbers, reflecting the additional support provided by repulsive self-interactions against gravitational collapse. The excited-state and unbound regions show a qualitatively consistent structure across the different initial ans\"atze explored, suggesting that they are robust features of the stationary GPP solution space.

We also investigated the dependence of the ground-state halo profiles on the particle properties and on the total particle number. Increasing the scattering length produces more extended and less centrally concentrated halos, while increasing either the boson mass or the total particle number leads to more compact configurations. These trends are consistent with the physical competition between quantum pressure, repulsive self-interactions, and self-gravity.

From the set of converged ground-state solutions we extracted empirical scaling relations for the characteristic radius $R_{99}$. In the weakly interacting regime, $a_{s} = 10^{-80} \textrm{ m}$, the fitted relation
\begin{equation}
    R_{99}\propto m_\phi^{-2.996}N^{-0.998}
\end{equation}
is consistent with the standard non-interacting Schr\"odinger--Poisson mass--radius scaling, $R\propto m_\phi^{-2}M_{\rm H}^{-1}$. When repulsive self-interactions are included, the radius acquires an explicit dependence on $a_s$. In particular, the scaling obtained at fixed particle number $N=10^{100}$,
\begin{equation}
    R_{99}\propto a_s^{0.481}m_\phi^{-1.545},
\end{equation}
resembles the Thomas--Fermi expectation
$R_{\rm TF}\propto a_s^{1/2}m_\phi^{-3/2}$, although the fitted exponents indicate that the solutions do not lie in the asymptotic Thomas--Fermi regime. Instead, they occupy an intermediate regime in which quantum pressure, self-interactions, and gravity all contribute to the equilibrium structure.
When we let all three parameters $m_{\phi}$, $N$ and $a_{s}$ to vary we obtain the fitting
\begin{equation}
    R_{99} \propto m_\phi^{-2.096}N^{-0.391}a_{s}^{0.308}.
\end{equation}
particularly for the analyzed parameter space.

Excited-state configurations were also found as stationary solutions of the nonlinear eigenvalue problem. These configurations exhibit radial nodes in the condensate wavefunction and oscillatory density shells. Although they are mathematically valid stationary branches of the GPP system, previous time-dependent studies indicate that excited states are generally less dynamically stable than the ground-state branch and tend to relax through gravitational cooling. In the present work they are therefore used primarily to characterize the structure of the stationary solution space.

As an astrophysical application, we used ground-state GPP configurations to model the rotation curves of the dwarf galaxies KK98-251 and UGC01281 from the SPARC sample. For $m_\phi=10^{-22}\,\mathrm{eV}$, the resulting halos reproduce the observed rotation curves with physically plausible halo masses and characteristic radii. In these examples, the ground-state condensate alone accounts for the dark matter contribution over the radial range probed by the data, without requiring an additional extended envelope.

Finally, we discussed the implications of Lyman-$\alpha$ forest constraints. Increasing the scattering length allows ground-state solutions to exist at boson masses compatible with current Lyman-$\alpha$ bounds, reaching $m_\phi\sim10^{-20}$--$10^{-18}\,\mathrm{eV}$ for $a_s\sim10^{-69}\,\mathrm{m}$. However, within the stationary halo configurations studied here, these high-mass solutions do not reproduce the dwarf-galaxy rotation curves considered in this work. Future work should extend the analysis to a larger sample of galaxies and perform systematic rotation-curve fits across a broader region of parameter space in order to determine whether this conclusion remains valid in a more general setting.

Overall, our results show that treating the total particle number $N$ as an explicit control parameter provides a useful way to map fundamental bosonic properties and the total particle content into macroscopic halo observables. The resulting phase diagrams and scaling relations offer a controlled framework for studying stationary self-gravitating Bose--Einstein condensates as dark matter halos. Future work should investigate the dynamical stability of the ground-state and excited-state branches, extend the analysis to configurations with an outer halo envelope, and perform systematic fits to larger samples of observed rotation curves.

\section*{Acknowledgments}
FG thanks the MCTP-UNACH and collaborating theoretical groups at UNAM for their support. JM thanks the program \textit{Investigadoras e Investigadores por México} of SECIHTI.

\bibliographystyle{unsrt}
\bibliography{gpp}

\appendix

\section{Rotation-curve modelling}\label{app:rotation_curves}

In this appendix we summarize the rotation-curve prescription used to compare the stationary Gross--Pitaevskii--Poisson halo solutions with observed galactic kinematics. For a spherically symmetric mass distribution, the circular velocity generated by the enclosed mass is
\begin{equation}
    v_c^2(r)
    =
    \frac{G M_{\rm enc}(r)}{r},
\end{equation}
where
\begin{equation}
    M_{\rm enc}(r)
    =
    4\pi
    \int_0^r
    r'^2\rho(r')\,dr' .
\end{equation}
For the GPP halo, the mass density is
\begin{equation}
    \rho_{\rm DM}(r)
    =
    m_\phi |\psi(r)|^2,
\end{equation}
so that the dark-matter contribution to the circular velocity is obtained directly from the stationary condensate profile.

The total rotation curve is modeled as the quadratic sum of the baryonic and dark-matter contributions,
\begin{equation}\label{eq:vc_2}
    v_{\rm tot}^2(r)
    =
    v_{\rm gas}^2(r)
    +
    \Upsilon_{\star}\,v_{\rm disk}^2(r)
    +
    \Upsilon_{\rm bulge}\,v_{\rm bulge}^2(r)
    +
    v_{\rm DM}^2(r),
\end{equation}
where \(v_{\rm gas}\), \(v_{\rm disk}\), and \(v_{\rm bulge}\) are the gas, stellar-disk, and bulge contributions provided by the observational rotation-curve data. The factors \(\Upsilon_{\star}\) and \(\Upsilon_{\rm bulge}\) denote the corresponding stellar mass-to-light ratios. In the applications presented in Sec.~\ref{sec:results}, we fix the stellar mass-to-light ratio to \(\Upsilon_{\star}=0.574\), while the dark-matter contribution is computed from the ground-state GPP solution.

Equivalently, the baryonic contribution may be written as
\begin{equation}
    v_b^2(r)
    =
    v_{\rm gas}^2(r)
    +
    \Upsilon_{\star}\,v_{\rm disk}^2(r)
    +
    \Upsilon_{\rm bulge}\,v_{\rm bulge}^2(r),
\end{equation}
so that
\begin{equation}
    v_{\rm tot}^2(r)
    =
    v_b^2(r)
    +
    v_{\rm DM}^2(r).
\end{equation}
This is the prescription used to obtain the rotation curves shown in Fig.~\ref{fig:gauss_galaxies}.

\section{Fitting with fix mass}
For fixed $m_{\phi}=10^{-22}\,\mathrm{eV}$ in the region $3\times10^{96}\leq N \leq 7\times10^{96}$ and $10^{-77} \textrm{ m}\leq a_{s}\leq 10^{-76} \textrm{ m}$, we find 
\begin{equation}
    R_{99} = R_{\square} \left( \frac{a_{s}} {10^{-77}\,\mathrm{m}} \right)^{0.336 \pm 0.001} \left( \frac{N} {10^{97}} \right)^{-0.326 \pm 0.003},
\end{equation}
with $R_{\square} = 2.751 \pm 0.008 \textrm{ kpc}$.

\end{document}